\documentstyle[12pt,epsfig,eqsecnum,preprint,floats,aps]{revtex}
\input FEYNMAN

\newcommand{\ie}{{\it i.e.}\ }
\newcommand{\eg}{{\it e.g.}\ }
\newcommand{\cf}{{\it cf.}\ }
\newcommand{\be}{\begin{eqnarray}}
\newcommand{\ee}{\end{eqnarray}}
\newcommand{\dslash}{\partial \!\!\! /}
\newcommand{\sgn}{{\rm sgn}}
\newcommand{\TTr}{{\bf Tr}}
\newcommand{\tr}{{\rm tr}}
\newcommand{\DDet}{{\bf Det}}
\newcommand{\A}{{\cal A}}
\renewcommand{\AA}{{\bf A}}
\newcommand{\BB}{{\bf B}}
\newcommand{\D}{{\cal D}}
\newcommand{\G}{{\cal G}}
\newcommand{\GG}{{\bf G}}
\newcommand{\OO}{{\bf O}}
\newcommand{\qq}{{\bf q}}
\newcommand{\PPsi}{{\mbox{\boldmath $\Psi$}}}
\newcommand{\xxi}{{\mbox{\boldmath $\xi$}}}
\renewcommand{\L}{{\cal L}}
\newcommand{\zr}[1]{\mbox{\hspace*{#1em}}}
\newcommand{\ID}{\mbox{{\sf 1}\zr{-0.16}\rule{0.04em}{1.55ex}\zr{0.1}}}

\begin{document} 
\draft
\date{\today}
\preprint{\parbox[t]{45mm}{UNITU-THEP-08/1996\\ nucl-th/9609012}}
\title{Baryons as Hybrids of Solitons and Three Quark Bound States}
\baselineskip=18 true pt
\author{U.\ Z\"uckert, R.\ Alkofer, H.\ Weigel, and H.\ Reinhardt}
\address{Institute for Theoretical Physics,
T\"ubingen University \\
Auf der Morgenstelle 14,
D-72076 T\"ubingen, Germany}
\maketitle

\begin{abstract}
A hybrid model for baryons based on a dynamical interplay between
relativistic three--quark bound states and soliton configurations of
mesons is constructed. The Bethe-Salpeter equation for diquarks and
the Faddeev equation for diquark-quark bound states in the background
of a soliton is solved. The results show that baryons are very much
like hybrids containing both, solitonic meson clouds and three--quark
correlations. 
\end{abstract}


\noindent



\section{Introduction}

At present there exist two generic classes of models to describe
baryons. On the one hand, there is the picture of baryons as chiral
solitons. The soliton picture is based on considering QCD for an
arbitrary number of color degrees of freedom, $N_C$. In the infinite
color limit, $N_C \to \infty$, QCD reduces to an effective theory of
infinitely many weakly interacting mesons \cite{tHo74}. Although this
effective meson theory cannot be constructed explicitly, Witten
conjectured that within this theory baryons emerge as solitons
\cite{Wit79}. Based on this conjecture phenomenological effective
meson theories have been developed which possess soliton
solutions. The most prominent is perhaps the Skyrme model
\cite{Sky61,Adk83,Hol93}. In the limit of an infinite number of colors
the soliton description is the only model for baryons. On the other
hand, for a finite number of colors a baryon is customarily considered
as a bound state of three valence quarks. Such valence quark models
are motivated by high energy scattering experiments which have
revealed a partonic substructure. Starting with these experimental
facts many models, which are based on the valence quark picture, have
been developed. These models include the non-relativistic quark models
\cite{Gel64a,Kar68,Fai68} and its relativistic extensions
\cite{Fey71}, parton models which are directly based on the scale
invariance \cite{Kog74}, bag models \cite{Has78} and diquark-quark
models \cite{Lic67,DeS67}. 

The valence quark picture directly leads to the quantum numbers
of a physical baryon whereas the soliton can only be interpreted as a
baryon with good spin and flavor quantum numbers after collective
quantization. Yet, baryons as solitons are conceptually better suited
for the description of low energy properties because they
straightforwardly embed the useful feature of chiral symmetry and its  
spontaneous breaking. In any event, despite their successes both
pictures possess only limited ranges of applicability. 

Since the advantages of both pictures are in some sense complementary
a unification of the two approaches seems desirable. In principle, the
chiral bag model \cite{Cho75,Rho94} represents such a combination
since inside the bag it contains explicit quark degrees of freedom
whereas a chiral soliton field surrounds the bag. As a consequence of
the Chesire cat principle \cite{Nad85} experimental measurable
quantities should not depend on the radius chosen for the bag. 
Explicit calculations show  that the Chesire cat principle is not
universally valid. In particular, the singlet axial matrix element
depends strongly on the bag radius \cite{Hos90,Hos93}. Therefore, a
hybrid model is desirable which connects both classes of models in a 
dynamical fashion. In this context the Nambu--Jona-Lasinio (NJL) model
\cite{Nam61} is for the moment unique. On the one hand, it possesses
soliton solutions \cite{Alk96a} of meson fields which themselves are
obtained as bound antiquark--quark states \cite{Ebe86}. On the other
hand, within this model baryons may be described as three--quark bound
states via the use of diquarks
\cite{Buc92,Ish93,Hua94,Mey94,Han95,Ish95,Buc95}, while the meson
fields are fixed to their vacuum expectation values. In particular
this model is unique because with the help of path integral
hadronization techniques \cite{Rei90} a consistent unification of both
approaches  is possible without any double counting of correlations. 

Because of the enormous computational effort needed, the realization
of a rigorous self-consistent solution is for the moment not
feasible. Nevertheless, an approximate evaluation of this hybrid
baryon can be accomplished within a four step procedure to be carried
out in this paper: In the proceeding section the transformation 
of the NJL model with a pointlike interaction of color octet current
into an effective theory of mesons, diquarks and baryons is described.
For completeness we briefly repeat the first two steps in the
beginning of the third section: First, we construct a static ground
state solution in the absence of diquark and elementary baryon
fields. Second, we solve the Bethe--Salpeter equation for a scalar
diquark in the solitonic background field \cite{Zue95a}. In addition,
we derive the Faddeev equation for arbitrary quark propagators. In
section \ref{Numeric} we discuss numerical results for the solutions
of the Bethe-Salpeter equation as well as of the Faddeev equation. In
section \ref{Hybrid} we employ these results to define the hybrid
model and discuss some static properties. We close with conclusions
and a outlook in section \ref{Conclusions}. Some details of the
calculation and a few lengthy formulas are left to appendices.  

\section{Hadronization of the NJL model}

As stated in the introduction we consider a NJL model for two flavors 
\be
\label{L_NJL}
 \L_{NJL} = \bar{q}\left(i\dslash-\hat{m}^0\right)q - \frac{1}{2} g
 j_{\mu}^a j^{\mu}_a
\ee
with a pointlike interaction of color octet flavor singlet currents
$j_{\mu}^a = \bar{q} \lambda_C^a \gamma_{\mu} q /2$. Here $q$ denotes
the quark spinors, $\hat{m}^0={\rm diag}(m_0^u,m_0^d)$ the current
quark mass matrix for two flavors and $\lambda_C^a (a=1,..,N_C^2-1)$
are the generators of color $SU(N_C)$. The interaction part can be
Fierz-rearranged to solely attractive channels 
\be
\L_{int}=g_1 \left(1-\frac{1}{N_C}\right) \left(\bar{q}\Lambda_\alpha q \right)
\left(\bar{q}\Lambda^\alpha q \right) +
\frac{g_2}{N_C} \left(\bar{q}\Gamma_\alpha q^c \right)
\left(\bar{q}^c\Gamma^\alpha q \right) ,
\ee
where $q^c=C\bar{q}^T$ denotes the charge conjugated Dirac spinor.
Furthermore, we have defined the vertices
\be
\Lambda^\alpha=\ID_c \frac{\tau^a}{2}{\cal O}^{\hbox{a}} , \quad
\Gamma^\alpha=\left(\frac{i\epsilon^A}{\sqrt{2}}\right)_c
\frac{\tau^a}{2}{\cal O}^{\hbox{a}} \quad \hbox{with} \quad
\alpha=\left(A,a,\hbox{a}\right)
\ee
for the quark-antiquark and the quark-quark interaction, respectively.
$\cal{O}^{\hbox{a}}$ corresponds to the set of Dirac matrices
\be
{\cal O}^{\hbox{a}} \in \{ \ID , i\gamma_5 ,
\frac{i\gamma^\mu}{\sqrt{2}} , \frac{i\gamma^\mu \gamma_5}{\sqrt{2}} \}
\ee
and $\tau^a$ are isospin matrices. Because of the Fierz-transformation
the coupling constants are restricted to $g_1=g_2=g$, which is
necessary to preserve consistent $N_C$ counting. Note, that the
diquark channel is suppressed by a factor $1/N_C$ in comparison to
the meson channel. Henceforth we will confine the discussion to the
physical case $N_C=3$, unless explicit noted.

To convert the pure quark NJL model (\ref{L_NJL}) into an effective
hadron theory\footnote{Similar hadronization approaches have been
considered in \cite{Cah89a,Ebe92}} \cite{Rei90} we introduce
collective meson $\varphi^\alpha$ and baryon fields $B_\alpha^\beta$
and $\bar{B}_\alpha^\beta$ into the generating functional
\be
 Z &=& \int \D q \D \bar q  \exp \left( \int \L_{NJL} \right)
\label{effTheo}
\ee
via the identities 
\be
\label{meson}
 \ID &=& \int \D \chi_\alpha \delta (\chi_\alpha -\bar{q}
 \Lambda_\alpha q)
   = \int \D \chi_\alpha \D \varphi^\alpha \exp \left(
   i \int \varphi^\alpha (\chi_\alpha -\bar{q} \Lambda_\alpha q)
   \right), \\
\label{baryon}
 \ID &=& \int \D \bar{B}_\alpha^\beta \D B_\alpha^\beta
 \delta \left( B_\alpha^\beta - \frac{2g_2}{3} (\bar{q}^c
 \Gamma_\alpha q) P^\beta q \right)
 \delta \left (\bar{B}_\alpha^\beta -\frac{2g_2}{3}\bar{q} P^\beta
 (\bar{q}  \Gamma_\alpha q^c ) \right)  \nonumber \\
 &=& \int \D \bar{B}_\alpha^\beta \D B_\alpha^\beta
 \D \bar{\Psi}^\alpha_\beta \D \Psi^\alpha_\beta \\ \nonumber
 && \qquad \exp \left( i \int \left[ \left\{ B_\alpha^\beta -
 \frac{2g_2}{3} \left(\bar{q}^c  \Gamma_\alpha q\right) P^\beta q \right\}
 \bar{\Psi}^\alpha_\beta  + \Psi^\alpha_\beta \left\{
 \bar{B}_\alpha^\beta - \frac{2g_2}{3} \bar{q} P^\beta \left(\bar{q}
 \Gamma_\alpha q^c \right) \right\} \right] \right).
\ee
Note, that we are working in Euclidean space where we have used 
the Wick rotation $t\rightarrow -i\tau$. For notational simplification
we use the abbreviation $\int = \int d^4x = \int d^3\mbox{\boldmath
$r$} d\tau$ in the exponents. The operator $P^\beta$ projects the
three quarks onto the quantum numbers of the considered physical
baryon states. In particular, the color of the third quark has to be
chosen to build a colorless baryon wavefunction.  

As intermediate building blocks for the baryon field we introduce also
diquark fields
\be
\label{diquark}
 \ID &=& \int \D \kappa_\alpha \D \kappa_\alpha^* \delta (\kappa_\alpha^*
 -\bar{q}\Gamma_\alpha q^c)\delta (\kappa_\alpha -\bar{q}^c
 \Gamma_\alpha q) \\ \nonumber
   &=& \int \D \kappa_\alpha \D \kappa_\alpha^* \D \Delta^\alpha \D
   \Delta^{\alpha *} \exp \left(
  \frac{i}{2} \int \left[ (\kappa_\alpha^* -\bar{q} \Gamma_\alpha
  q^c) \Delta^\alpha + \Delta^{\alpha *} (\kappa_\alpha -\bar{q}^c
  \Gamma_\alpha q)\right] \right) .
\ee
Replacing all terms of fourth order in the quark fields with help of
the constraints (\ref{meson}) and (\ref{diquark}), and integrating out
the auxiliary fields $\chi_\alpha, \kappa_\alpha$ and
$\kappa_\alpha^*$, one obtains the following form for the generating
functional 
\be
\label{effMod}
 Z &\sim& \int \D q \D \bar q \D \varphi \D \Delta \D \tilde{\Delta}
 \D \bar{B} \D B \D \bar{\Psi}^\alpha \D \Psi^\alpha \\
 &&  \exp \left( \int \left[ \bar{q} \left\{i \gamma^\mu
 \partial_\mu -m^0 -\varphi + \frac{4g_2}{3} \bar{\Psi}^\alpha
 \Psi^\alpha \right\} q
 +  \bar{q}\Delta^*_\alpha \Psi^\alpha +
 \bar{\Psi}^\alpha \Delta_\alpha q \right] \right) \nonumber \\
 &&  *\exp \left( - \int \left[ \frac{1}{2} \bar{q} \Delta q^c +
 \frac{1}{2} \bar{q}^c \tilde{\Delta} q + \frac{1}{8g_1} \tr \varphi^2 +
 \frac{3}{8g_2} \tr \tilde{\Delta} \Delta - i \left\{ \bar{\Psi}^\alpha_\beta
 B_\alpha^\beta + \bar{B}_\alpha^\beta \Psi^\alpha_\beta \right\}
 \right] \right), \nonumber
\ee
where we have introduced the compact matrix notation for the meson and
diquark fields $\varphi = \varphi_\alpha \Lambda^\alpha , \Delta =
\Gamma_\alpha \Delta^\alpha , \tilde{\Delta}= \Gamma_\alpha
\Delta^{\alpha*}$ as well as for the baryon sources $\Psi^\alpha =
\Psi^\alpha_\beta P^\beta$ and $ \bar{\Psi}^\alpha = 
P^\beta \bar{\Psi}^\alpha_\beta $. The symbol $\tr$ corresponds
to the trace over color, flavor and Dirac spinor degrees of freedom. 

To eliminate the quark degrees of freedom we are working within the
Nambu-Gorkov formalism developed originally in the theory of
superconductivity \cite{Nam60,Gor58,Sch64}. For that purpose we
introduce combined Grassmann fields\footnote{Quantities in the
Nambu-Gorkov formalism are denoted by boldface letters.} 
 \be
\label{kombGrassman}
 \qq = \left( \begin{array}{cc} q \\ \bar{q}^T \end{array} \right)
 \quad , \quad
 \bar{\qq} = \left( \bar{q} \ ,\ q^T \right)
\ee
for the quark fields and 
\be
 \PPsi^\alpha = \left( \begin{array}{c} \Psi^{\alpha} \\
         \bar{\Psi}^{\alpha T}  \end{array} \right)
 \quad , \quad
 \bar{\PPsi}^\alpha = \left( \bar{\Psi}^\alpha , \Psi^{\alpha T} \right)
\ee
for the baryon sources. Now, we integrate out the quark fields
$\qq$ and $\bar{\qq}$ with help of the Nambu-Gorkov formula
\be
\label{Nambu-Gorkov}
 && \int \D q \D \bar q \exp \left( \bar{q} \G^{-1} q  -
 \frac{1}{2} \bar{q} \Delta q^c - \frac{1}{2} \bar{q}^c \tilde{\Delta}
 q  + \left[ \bar{q} \Delta^{* \alpha} \Psi_\alpha +
 \bar{\Psi}_\alpha \Delta^\alpha q \right] \right) \nonumber \\
 &=& \int \D \qq \D \bar{\qq} \exp \left( \frac{1}{2}
 \left[ \bar{\qq} \GG^{-1} \qq + \bar{\qq} \xxi + \bar{\xxi} \qq
 \right] \right) = \left( \DDet \GG^{-1} \right)^{\frac{1}{2}} \exp
 \left( \frac{1}{2} \bar{\xxi} \GG \xxi \right)  ,
\ee
where we have defined
\be
\label{xxi}
 \xxi = \left( \begin{array}{cc} \Delta^{* \alpha} \Psi_\alpha \\
 -\bar{\Psi}^T_\alpha \Delta^\alpha \end{array} \right)
  \quad , \quad
 \bar{\xxi} = \left( \bar{\Psi}_\alpha \Delta^\alpha \ , \
 -\Delta^{* \alpha} \Psi^T_\alpha \right) ,
\ee
and the inverse quark Greens function 
\be
\label{green}
 \GG^{-1} = \left(\begin{array}{cc}
               \G^{-1}           & -\Delta C \\
               -C \Delta^* & -V\tilde{\G}^{-1}V^\dagger
            \end{array} \right) ,
\qquad \tilde{\G} = V^\dagger \G^T V .
\ee
The off-diagonal elements of $\GG$ are the so-called anomalous
Greens functions which are related to the amplitude for adding or
subtracting a pair of quarks to the system. The transformation
operator $V=JG$, which we have introduced for technical reasons, is a
combination of the self-adjoint unitary transformation
$J=i\beta\gamma_5$ and the G-parity operator $G=e^{i\pi\tau_2/2}C$. 

The normal quark Greens function is represented by
\be
\label{Quarkprop}
 {\G}^{-1} = i \gamma^\mu \partial_\mu - \varphi + \frac{4g_2}{3}
 \bar{\Psi}^\alpha  \Psi^\alpha = {\G}^{-1}_0 + \frac{4g_2}{3}
 \bar{\Psi}^\alpha  \Psi^\alpha .
\ee
From eq.~(\ref{xxi}) we observe that the baryon sources $\Psi^\alpha$
are contracted with the diquark fields $\Delta_\alpha$. Hence these
sources only couple to a single quark. Since in the ladder
approximation (described below) a three quark bound state cannot be
affected by such a coupling it is sufficient to restrict the quark 
Greens function to ${\G}^{-1}_0$.   

Finally, the generating functional is given by
\be
\label{gener_func}
 Z[B_\alpha^\beta,\bar{B}_\alpha^\beta] &\sim& \int \D \varphi \D \Delta
 \D \tilde{\Delta} \D \bar{\Psi}^\alpha \D \Psi^\alpha \nonumber \\
 &&\qquad \exp\left(\A[\varphi, \Delta, \tilde{\Delta}, \bar{\Psi}^\alpha,
 \Psi^\alpha] + i \left( \bar{\Psi}^\alpha_\beta B_\alpha^\beta +
 \bar{B}_\alpha^\beta \Psi^\alpha_\beta \right) \right)
\ee
with the effective action 
\be
\label{effWirk}
 \A[\varphi, \Delta, \tilde{\Delta},\bar{\Psi}^\alpha, \Psi^\alpha ]
 &=& \\ \nonumber
 && \A_q[\varphi, \Delta,  \tilde{\Delta}, \bar{\Psi}^\alpha,
 \Psi^\alpha]  + \A_{val}[\varphi, \Delta, \tilde{\Delta},
 \bar{\Psi}^\alpha, \Psi^\alpha ]
 + \A_m[\varphi] +  \A_{d}[\Delta, \tilde{\Delta}].
\ee
The first part, the so-called quark determinant, 
\be
\label{fermdet}
 \A_q = \frac{1}{2} \TTr \log \GG^{-1} 
\ee
carries the full information of the underlying quark spectrum via the
functional trace $\TTr$. The second part 
\be
\label{A_b^2}
 \A_{val} = \frac{1}{2} \bar{\PPsi}^\alpha
 (\GG^{val}_{B})_{\alpha \beta} \PPsi^\beta 
\ee
contains the valence quark part of the baryon propagator
$(\GG^{val}_B)_{\alpha \beta}$ originating from the inversion of the
quark propagator $\GG_0$ (\cf appendix B for the derivation of
$(\GG^{val}_B)_{\alpha \beta}$.).

Both residual terms 
\be
\label{A_m}
 \A_m &=& -\frac{1}{8g_1}\int d^4x \left[\tr\varphi^2\right] ,\nonumber \\
 \A_d &=& -\frac{3}{8g_2}\int d^4x \left[\tr\tilde{\Delta} \Delta
 \right]
\ee
are pure mass terms for the meson and diquark fields,
respectively\footnote{Note the additional factor $N_C=3$ in the diquark
mass term $\A_d$ as compared to the mass term $\A_m$ of the mesons.}.

\newpage

\section{Two and three quark correlations in a solitonic background} 

\subsection{The static soliton background}

In this section we are interested in the behavior of quark-quark and
diquark-quark bound states in the background field of a soliton
configuration. In the first step we neglect the explicit baryon
sources. With this restriction it is possible to expand the effective
action (\ref{effWirk})  
\be
 \A[\varphi, \Delta, \tilde{\Delta},\bar{\Psi}^\alpha=0,
 \Psi^\alpha=0]  = \A^{(0)}[\varphi]  +\A^{(1)}[\varphi,\Delta,
 \tilde{\Delta}] +\A^{(2)}[\varphi, \Delta, \tilde{\Delta}] + \ldots  
\ee
up to second order in the diquark fields where (see also
eqs.~(\ref{A_b^2}) and (\ref{A_m}))
\be
\label{expand}
 \A^{(0)}[\varphi] &=& \A_q^{(0)} + \A_{val} + \A_m , \\
 \A^{(1)}[\varphi,\Delta,\tilde{\Delta}] &=& \A_q^{(1)} ,\\
 \A^{(2)}[\varphi,\Delta,\tilde{\Delta}] &=& \A_q^{(2)} + \A_d .
\ee
Within Schwinger's proper time regularization description\cite{Sch51}
the parts stemming from the quark determinant (\ref{fermdet}) can be
written as
\be
\label{A_q}
 \A_q^{(0)} &=& -\frac{1}{4} \TTr \int_{1/\Lambda^2}^{\infty}
 \frac{ds}{s} e^{-s\AA_0} , \nonumber \\
\label{A1}
 \A_q^{(1)} &=& \frac{1}{4} \TTr \int_{1/\Lambda^2}^{\infty} ds \AA_1
 e^{-s\AA_0} ,\\
 \A_q^{(2)} &=& \frac{1}{4} \TTr \int_{1/\Lambda^2}^{\infty} ds \AA_2
 e^{-s\AA_0} -\frac{1}{4} \TTr \int_{1/\Lambda^2}^{\infty} ds s
 \int_0^1 d\alpha \alpha \AA_1 e^{-s\alpha\AA_0} \AA_1
 e^{-s(1-\alpha)\AA_0} \nonumber .
\ee
The Nambu-Gorkov matrices $\AA_i$ originating from the expansion of the
operator $(\GG^{-1})^\dagger \GG^{-1} = \AA_0 + \AA_1 + \AA_2$ have
the explicit form
\be
 \AA_0 &=& \left( \begin{array}{cc}
     (\G^{-1})^\dagger \G^{-1}  & 0 \\
       0          & V^{-1} (\tilde{\G}^{-1})^\dagger \tilde{\G}^{-1} V
                  \end{array} \right) , \\
 \AA_1 &=& \left( \begin{array}{cc}
     0  & -(\G^{-1})^\dagger \Delta C - \tilde{\Delta}^\dagger C V^{-1}
     \tilde{\G}^{-1} V \\
   C \Delta^\dagger \G^{-1} + V^{-1} (\tilde{\G}^{-1})^\dagger V C
   \tilde{\Delta} &  0
                  \end{array} \right) , \\
 \AA_2 &=& \left( \begin{array}{cc}
     \tilde{\Delta}^\dagger \tilde{\Delta}  & 0 \\
       0          & -C \Delta^\dagger \Delta C
                  \end{array} \right) .
\ee

By expanding $\A_q^{(0)}$ to quadratic order in $\varphi$ we can
relate \cite{Ebe86,Wei92} the parameters ($g_1$, $m_0=m_0^u=m_0^d$ and 
$\Lambda$) of the model to the pion mass $m_\pi=135$MeV and the pion
decay constant $f_\pi=93$MeV, leaving undetermined just one parameter
which we choose to be the constituent quark mass $m$. This mass is the
vacuum expectation value of the scalar field and reflects the
spontaneous breaking of chiral symmetry. In principle the diquark
coupling constant $g_2$ is fixed by the Fierz transformation, however,
we will relax the restriction $g_2/g_1=1$ thus effectively treating
$g_2/g_1$ as a parameter to study the influence of the diquarks.  

Note, that for each term in eq.~(\ref{A_q}) the Nambu-Gorkov trace can
be worked out analytically. In addition, due to the diagonality of 
$\AA_0$ and the vanishing trace of $\AA_1$ it is obvious that 
$\Delta=\tilde{\Delta}=0$ is always a solution of the equation of motion
though it is not necessarily the solution of least action. For the
remaining functional trace we use the color degenerated eigenstates of
the inverse propagator for quarks in the background of a static meson
field $\varphi$ 
\be
\label{propagator}
 {\G}^{-1}|n , \nu\rangle = \left(i \dslash - \varphi\right)
 |n , \nu\rangle = \beta \left( - \partial_\tau - h_\Theta\right)
 |n , \nu\rangle = \beta \left( i \omega_n - \epsilon_{\nu}\right) |n ,
 \nu \rangle 
\ee
and perform an additional color trace $\tr_C$. In analogy, the charge
conjugated quark propagator $\tilde{\G}^{-1}_0$ can be worked out similarly:
\be
 \tilde{\G}^{-1}_0|n , \nu\rangle = V^\dagger \left(i \dslash -
 \varphi\right)^T V |n , \nu\rangle = \beta \left( \partial_\tau -
 h_\Theta\right) |n , \nu\rangle = \beta \left( - i \omega_n -
 \epsilon_{\nu}\right) |n,  \nu \rangle .
\ee

For the meson fields we make use of the polar decomposition 
\be
\label{chiral}
 \varphi = \Phi\ U^{\gamma_5} = m\ U^{\gamma_5},
\ee
where we have fixed the chiral radius $\Phi$ to its vacuum expectation
value $m$. Additionally, we adopt the well-known hedgehog ansatz for
the chiral field  
\be
\label{chiral-field}
 U = {\rm exp}\Big(i{\mbox{\boldmath $\tau$}}
 \cdot{\hat{\mbox {\boldmath $r $}} }\Theta(r)\Big).
\ee
Due to the static field configuration it is possible to extract from
the zero order term $\A^{(0)}=-TE_{sol}[\Theta]$ (\ref{expand}) an
energy functional  
\be
\label{ESol}
 E_{sol}[\Theta] &=& E_{sea} + E_{val} \\
 E_{sea} &=& \frac{3}{2} \int_{1/\Lambda^2}
  ^\infty ds \frac{1}{\sqrt{4 \pi s^{3}}} \sum_{\mu} e^{-s \epsilon_\mu^2} 
  + m_\pi^2f_\pi^2\int d^3r \big(1-{\rm cos}\Theta(r)\big) \nonumber \\
  E_{val} &=& \frac{3}{2}\epsilon_{val}\big( 1+\sgn(\epsilon_{val})
  \big)  . \nonumber 
\ee
Neglecting quark-quark correlations, the valence quark
contribution $E_{val}$ is composed of the color degenerated
single particle energies of the valence quark level. This state is the
only one which is bound by the soliton, \ie $-m<\epsilon_{val}<m$. The
soliton configuration is characterized by the chiral angle
$\Theta_{sc}(r)$ which is self-consistently determined by extremizing
the energy functional $E[\Theta]$ \cite{Rei88a,Mei89,Alk90a}.   

\subsection{Bethe-Salpeter equation for diquarks}

In the following we derive the equation of motion for the diquark
fields in the solitonic meson background field. The solutions of this
equation define the eigenmodes $b_\alpha$ and $b^*_\alpha$ of the
diquark fields 
\be
 \Delta &=& b_\alpha (\Delta(\mbox {\boldmath $r $},t)\Gamma C
 V^{-1})^\alpha =  b_\alpha \Gamma_{diq}^\alpha(\mbox {\boldmath $r
 $},t),  \nonumber \\  \tilde{\Delta} &=& (V C \Gamma \Delta(\mbox
 {\boldmath $r $},t))^\alpha b^*_\alpha = 
 b^*_\alpha  \tilde{\Gamma}_{diq}^\alpha(\mbox {\boldmath $r $},t)
\ee
which we have redefined including the matrices $V$ and $C$ into the
diquark vertices. For the present qualitative discussion it is
sufficient to consider only an S-wave scalar diquark field. In respect
of the Pauli principle the only possible ansatz for the vertex functions
$\Gamma_{diq}^\alpha(\mbox {\boldmath $r $},t)$ is 
\be
 \Gamma_{diq}^\alpha(\mbox {\boldmath $r $},t) = \Delta(r,t) \Gamma^\alpha
 \quad , \quad  \tilde{\Gamma}_{diq}^\alpha(\mbox {\boldmath $r $},t)
 = \Delta^*(r,t) \Gamma^\alpha \quad  , \quad
 \Gamma^\alpha = - \frac{\lambda^a_C}{\sqrt{2}} \frac{\tau_2}{2} i
 \gamma_5
\ee
with $\lambda^a_C (a=2,5,7)$ being the antisymmetric Gell-Mann
matrices of the color group.

Using the Fourier-transform in the time coordinate
\be
 \Delta(r,\tau) = \int \frac{d\omega}{2\pi} \Delta(r,i\omega)
 e^{-i\omega \tau} 
\ee
and the spatial representation $\Psi_\nu(\mbox
{\boldmath $r $}) = \langle \mbox {\boldmath $r $} | \nu \rangle $ for
the quark eigenstates the residual trace can be worked out. The second
order term of the effective action, $\A_q^{(2)}$, together with $\A_d$
determines the Bethe-Salpeter equation as the equation of motion for
the diquark fields $\delta\A^{(2)}/\delta\Delta^*(r,-i\omega)=0$:   
\be
\label{BS}
 \left. r^{2} \left[ \int dr^\prime r^{\prime 2}
 D^{-1}_{\rm{diq}}(r,r^\prime  ;i\omega) \Delta_\alpha(r^\prime,i\omega)
 \right] \right|_{\omega^2=-\omega_{diq}^2}  = 0 .
\ee
The inverse diquark propagator 
\be
 D^{-1}_{\rm{diq}}(r,r^\prime ;i\omega) =  
 K(r,r^\prime;i\omega) - \frac{\pi m_\pi^2 f_\pi^2}{2m_0m} 
 \delta(r - r^\prime) 
\ee
is expressed in terms of a local mass term and a bilocal kernel 
\be
 \label{kernel}
K(r^\prime,r;i\omega) = \sum_{\nu \mu} \int d\Omega d\Omega^\prime
R(i\omega;\epsilon_{\nu},\epsilon_{\mu}) \left[
\Psi_\nu^\dagger(\mbox {\boldmath $r^\prime $})
\Psi_\mu(\mbox {\boldmath $r^\prime$}) \Psi_\mu^\dagger(\mbox {\boldmath $r $})
\Psi_\nu(\mbox {\boldmath $r $}) \right] 
\ee
where
\be
 \label{regfunc}
R(i\omega;\epsilon_{\nu},\epsilon_{\mu}) &=& \frac{1}{16 \sqrt{\pi}} \left[
\frac{1}{2 | \epsilon_{\nu} |}
\Gamma\left(\frac{1}{2},\frac{\epsilon_\nu^2}{\Lambda^2}\right)
 + \frac{1}{2 | \epsilon_{\mu} |}
\Gamma\left(\frac{1}{2},\frac{\epsilon_\mu^2}{\Lambda^2}\right)
 \right. \\*
&-& \left. \left(\omega^2 + \left( \epsilon_{\nu} -
\epsilon_{\mu} \right)^2 \right) \int_0^1 d\alpha \alpha
\frac{\Gamma\left(\frac{3}{2},\left[\alpha \epsilon_\mu^2 + (1-\alpha)
\epsilon_\nu^2 +\alpha(1-\alpha)\omega^2 / \Lambda^2 \right] \right)}
{\left[\alpha \epsilon_\mu^2 + (1-\alpha) \epsilon_\nu^2
+\alpha(1-\alpha)\omega^2 \right]^{3/2}} \right] \nonumber
\ee
in the proper-time regularized version of the quark loop. 

For the ongoing discussion a normalization of the diquark field
$\Delta_\alpha(r,i\omega)$ is necessary. In principle such a
normalization would be obtained in the framework of second
quantization. Equivalently, we demand that the total baryon charge, to
which the diquarks contribute, equals unity. For that purpose we first
calculate the expectation value of the baryon number operator
$\hat{B}=\ID_c/N_c=\ID_c/3$ (see appendix C for details.). In analogy
to the effective action (\ref{effWirk}) the expression (\ref{op}) is
expanded in terms of diquark fields. The leading order yields the
value $1/3$, which is nothing but the contribution of a single
quark. Therefore, we require for the second order term the
normalization condition  
\be
\label{norm}
 B_{diq} \stackrel{!}{=} \frac{2}{3}
\ee
with $B_{diq}$ defined in (\ref{B_diq}). As the solutions of the
Bethe-Salpeter equation appear always in pairs $\pm i\omega_{diq}$,
the normalization condition (\ref{norm}) allows us to distinguish 
between diquark and antidiquark solutions.  

\subsection{Faddeev equation for a quark-diquark bound state}

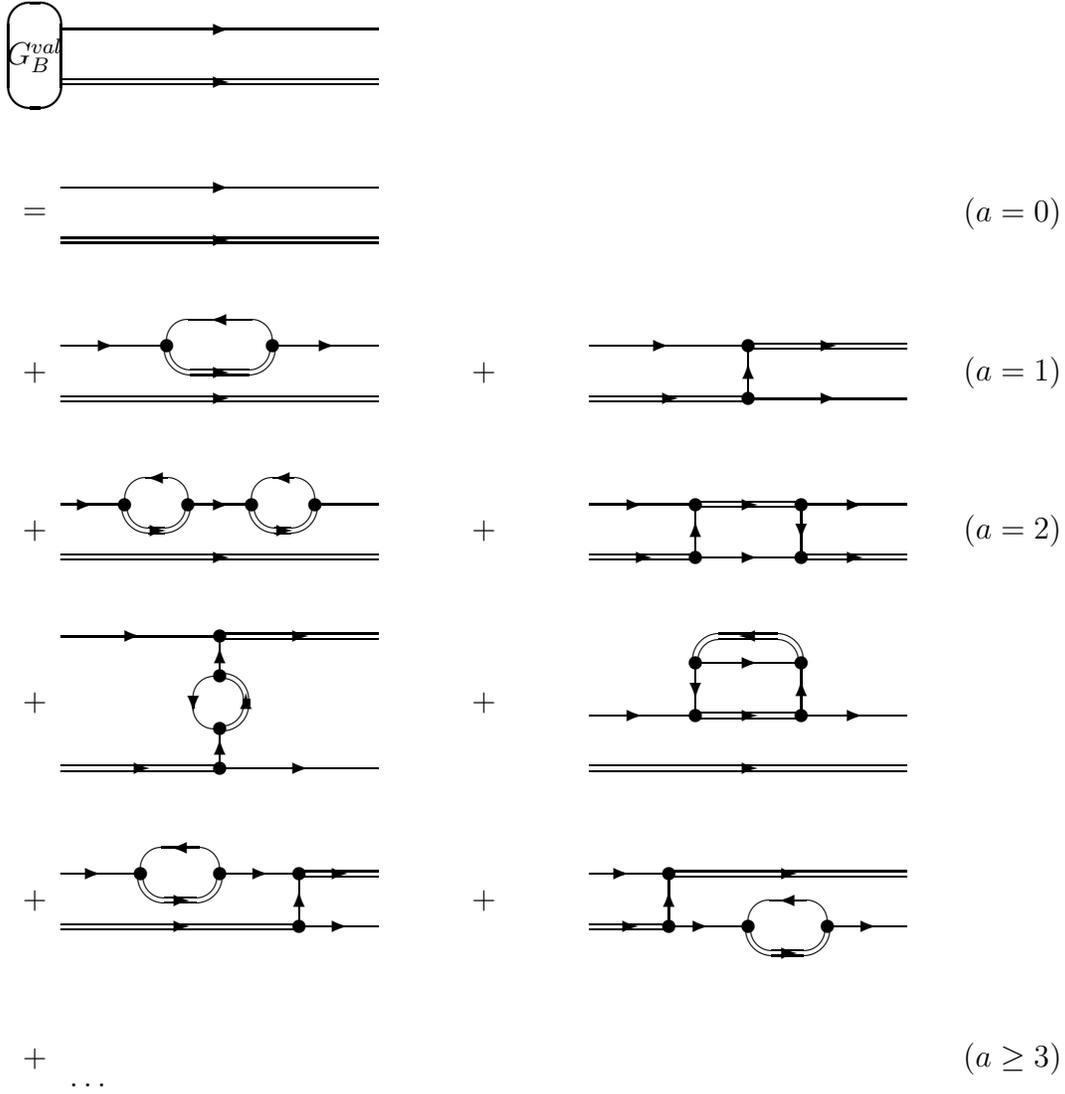
\begin{figure}
\vspace{1cm}
\centerline{
\begin{picture}(40000,43000)
\put(0,40000)
{\begin{picture}(25000,4000)
\thicklines
\put(1000,2000){\oval(2000,4000)}
\put(1000,2000){\makebox(0,0){$G^{val}_B$}}
\thinlines
\drawline\fermion[\E\REG](2000,3000)[12000]
\thicklines
\drawarrow[\LDIR\ATBASE](\pmidx,\pmidy)
\thinlines
\global\advance\fermionfronty by -1900
\drawline\fermion[\E\REG](\fermionfrontx,\fermionfronty)[\fermionlength]
\global\advance\fermionfronty by -200
\drawline\fermion[\E\REG](\fermionfrontx,\fermionfronty)[\fermionlength]
\global\advance\pmidy by 100
\thicklines
\drawarrow[\LDIR\ATBASE](\pmidx,\pmidy)
\end{picture}}
\put(0,34000)
{\begin{picture}(40000,4000)
\put(1000,2000){\makebox(0,0){$=$}}
\thinlines
\drawline\fermion[\E\REG](2000,3000)[12000]
\thicklines
\drawarrow[\LDIR\ATBASE](\pmidx,\pmidy)
\thinlines
\global\advance\fermionfronty by -1900
\drawline\fermion[\E\REG](\fermionfrontx,\fermionfronty)[\fermionlength]
\global\advance\fermionfronty by -200
\drawline\fermion[\E\REG](\fermionfrontx,\fermionfronty)[\fermionlength]
\global\advance\pmidy by 100
\thicklines
\drawarrow[\LDIR\ATBASE](\pmidx,\pmidy)
\put(38000,2000){\makebox(0,0){$(a=0)$}}
\end{picture}}
\put(0,28000)
{\begin{picture}(40000,4000)
\put(1000,2000){\makebox(0,0){$+$}}
\thinlines
\drawline\fermion[\E\REG](2000,3000)[4000]
\put(\fermionbackx,\fermionbacky){\circle*{500}}
\thicklines
\drawarrow[\LDIR\ATTIP](\pmidx,\pmidy)
\thinlines
\global\advance\fermionbackx by 2000
\put(\fermionbackx,\fermionbacky){\oval(4000,2000)[t]}
\put(\fermionbackx,\fermionbacky){\oval(3800,1800)[b]}
\put(\fermionbackx,\fermionbacky){\oval(4200,2200)[b]}
\thicklines
\global\advance\fermionbacky by 1000
\drawarrow[\W\ATBASE](\fermionbackx,\fermionbacky)
\global\advance\fermionbacky by -2000
\drawarrow[\E\ATBASE](\fermionbackx,\fermionbacky)
\global\advance\fermionbackx by 2000
\global\advance\fermionbacky by 1000
\thinlines
\put(\fermionbackx,\fermionbacky){\circle*{500}}
\drawline\fermion[\E\REG](\fermionbackx,\fermionbacky)[4000]
\thicklines
\drawarrow[\LDIR\ATBASE](\pmidx,\pmidy)
\thinlines
\drawline\fermion[\E\REG](2000,1100)[12000]
\global\advance\fermionfronty by -200
\drawline\fermion[\E\REG](\fermionfrontx,\fermionfronty)[\fermionlength]
\global\advance\pmidy by 100
\thicklines
\drawarrow[\LDIR\ATBASE](\pmidx,\pmidy)
\put(18000,2000){\makebox(0,0){$+$}}
\thinlines
\drawline\fermion[\E\REG](22000,3000)[6000]
\put(\fermionbackx,\fermionbacky){\circle*{500}}
\thicklines
\drawarrow[\LDIR\ATTIP](\pmidx,\pmidy)
\thinlines
\drawline\fermion[\E\REG](\fermionbackx,3100)[6000]
\global\advance\fermionfronty by -200
\drawline\fermion[\E\REG](\fermionfrontx,\fermionfronty)[\fermionlength]
\global\advance\pmidy by 100
\thicklines
\drawarrow[\LDIR\ATBASE](\pmidx,\pmidy)
\thinlines
\drawline\fermion[\E\REG](22000,1100)[6000]
\global\advance\fermionfronty by -200
\drawline\fermion[\E\REG](\fermionfrontx,\fermionfronty)[\fermionlength]
\global\advance\pmidy by 100
\thicklines
\drawarrow[\LDIR\ATBASE](\pmidx,\pmidy)
\global\advance\fermionbacky by 100
\put(\fermionbackx,\fermionbacky){\circle*{500}}
\thinlines
\drawline\fermion[\E\REG](\fermionbackx,\fermionbacky)[\fermionlength]
\thicklines
\drawarrow[\LDIR\ATBASE](\pmidx,\pmidy)
\thinlines
\drawline\fermion[\N\REG](\fermionfrontx,\fermionfronty)[2000]
\thicklines
\drawarrow[\LDIR\ATBASE](\pmidx,\pmidy)
\put(38000,2000){\makebox(0,0){$(a=1)$}}
\end{picture}}
\put(0,22000)
{\begin{picture}(40000,4000)
\put(1000,2000){\makebox(0,0){$+$}}
\thinlines
\drawline\fermion[\E\REG](2000,3000)[2400]
\put(\fermionbackx,\fermionbacky){\circle*{500}}
\thicklines
\drawarrow[\LDIR\ATTIP](\pmidx,\pmidy)
\thinlines
\global\advance\fermionbackx by 1200
\put(\fermionbackx,\fermionbacky){\oval(2400,2000)[t]}
\put(\fermionbackx,\fermionbacky){\oval(2200,1800)[b]}
\put(\fermionbackx,\fermionbacky){\oval(2600,2200)[b]}
\thicklines
\global\advance\fermionbacky by 1000
\drawarrow[\W\ATBASE](\fermionbackx,\fermionbacky)
\global\advance\fermionbacky by -2000
\drawarrow[\E\ATBASE](\fermionbackx,\fermionbacky)
\global\advance\fermionbackx by 1200
\global\advance\fermionbacky by 1000
\thinlines
\put(\fermionbackx,\fermionbacky){\circle*{500}}
\drawline\fermion[\E\REG](\fermionbackx,\fermionbacky)[2400]
\put(\fermionbackx,\fermionbacky){\circle*{500}}
\thicklines
\drawarrow[\LDIR\ATBASE](\pmidx,\pmidy)
\thinlines
\global\advance\fermionbackx by 1200
\put(\fermionbackx,\fermionbacky){\oval(2400,2000)[t]}
\put(\fermionbackx,\fermionbacky){\oval(2200,1800)[b]}
\put(\fermionbackx,\fermionbacky){\oval(2600,2200)[b]}
\thicklines
\global\advance\fermionbacky by 1000
\drawarrow[\W\ATBASE](\fermionbackx,\fermionbacky)
\global\advance\fermionbacky by -2000
\drawarrow[\E\ATBASE](\fermionbackx,\fermionbacky)
\global\advance\fermionbackx by 1200
\global\advance\fermionbacky by 1000
\thinlines
\put(\fermionbackx,\fermionbacky){\circle*{500}}
\drawline\fermion[\E\REG](\fermionbackx,\fermionbacky)[2400]
\drawline\fermion[\E\REG](2000,1100)[12000]
\global\advance\fermionfronty by -200
\drawline\fermion[\E\REG](\fermionfrontx,\fermionfronty)[\fermionlength]
\global\advance\pmidy by 100
\thicklines
\drawarrow[\LDIR\ATBASE](\pmidx,\pmidy)
\put(18000,2000){\makebox(0,0){$+$}}
\thinlines
\drawline\fermion[\E\REG](22000,3000)[4000]
\put(\fermionbackx,\fermionbacky){\circle*{500}}
\thicklines
\drawarrow[\LDIR\ATTIP](\pmidx,\pmidy)
\thinlines
\drawline\fermion[\E\REG](\fermionbackx,3100)[4000]
\global\advance\fermionfronty by -200
\drawline\fermion[\E\REG](\fermionfrontx,\fermionfronty)[\fermionlength]
\global\advance\pmidy by 100
\thicklines
\drawarrow[\LDIR\ATBASE](\pmidx,\pmidy)
\global\advance\fermionbacky by 100
\put(\fermionbackx,\fermionbacky){\circle*{500}}
\thinlines
\drawline\fermion[\E\REG](\fermionbackx,\fermionbacky)[4000]
\thicklines
\drawarrow[\LDIR\ATBASE](\pmidx,\pmidy)
\thinlines
\drawline\fermion[\E\REG](22000,1100)[4000]
\global\advance\fermionfronty by -200
\drawline\fermion[\E\REG](\fermionfrontx,\fermionfronty)[\fermionlength]
\global\advance\pmidy by 100
\thicklines
\drawarrow[\LDIR\ATBASE](\pmidx,\pmidy)
\global\advance\fermionbacky by 100
\put(\fermionbackx,\fermionbacky){\circle*{500}}
\thinlines
\drawline\fermion[\N\REG](\fermionbackx,\fermionbacky)[2000]
\thicklines
\drawarrow[\LDIR\ATBASE](\pmidx,\pmidy)
\thinlines
\drawline\fermion[\E\REG](\fermionfrontx,\fermionfronty)[4000]
\thicklines
\drawarrow[\LDIR\ATBASE](\pmidx,\pmidy)
\put(\fermionbackx,\fermionbacky){\circle*{500}}
\thinlines
\drawline\fermion[\N\REG](\fermionbackx,\fermionbacky)[2000]
\thicklines
\drawarrow[\S\ATBASE](\pmidx,\pmidy)
\global\advance\fermionfronty by 100
\thinlines
\drawline\fermion[\E\REG](\fermionfrontx,\fermionfronty)[4000]
\global\advance\fermionfronty by -200
\drawline\fermion[\E\REG](\fermionfrontx,\fermionfronty)[\fermionlength]
\global\advance\pmidy by 100
\thicklines
\drawarrow[\LDIR\ATBASE](\pmidx,\pmidy)
\put(38000,2000){\makebox(0,0){$(a=2)$}}
\end{picture}}
\put(0,14000)
{\begin{picture}(40000,8000)
\put(1000,3500){\makebox(0,0){$+$}}
\thinlines
\drawline\fermion[\E\REG](2000,6000)[6000]
\put(\fermionbackx,\fermionbacky){\circle*{500}}
\thicklines
\drawarrow[\LDIR\ATTIP](\pmidx,\pmidy)
\thinlines
\global\advance\fermionbacky by 100
\drawline\fermion[\E\REG](\fermionbackx,\fermionbacky)[6000]
\global\advance\fermionfronty by -200
\drawline\fermion[\E\REG](\fermionfrontx,\fermionfronty)[\fermionlength]
\global\advance\pmidy by 100
\thicklines
\drawarrow[\LDIR\ATBASE](\pmidx,\pmidy)
\thinlines
\drawline\fermion[\E\REG](2000,1100)[6000]
\global\advance\fermionfronty by -200
\drawline\fermion[\E\REG](\fermionfrontx,\fermionfronty)[\fermionlength]
\global\advance\pmidy by 100
\thicklines
\drawarrow[\LDIR\ATBASE](\pmidx,\pmidy)
\global\advance\fermionbacky by 100
\put(\fermionbackx,\fermionbacky){\circle*{500}}
\thinlines
\drawline\fermion[\E\REG](\fermionbackx,\fermionbacky)[6000]
\thicklines
\drawarrow[\LDIR\ATBASE](\pmidx,\pmidy)
\thinlines
\drawline\fermion[\N\REG](\fermionfrontx,\fermionfronty)[1500]
\thicklines
\drawarrow[\LDIR\ATBASE](\pmidx,\pmidy)
\put(\fermionbackx,\fermionbacky){\circle*{500}}
\thinlines
\global\advance\fermionbacky by 1000
\put(\fermionbackx,\fermionbacky){\oval(2000,2000)[l]}
\put(\fermionbackx,\fermionbacky){\oval(1800,1800)[r]}
\put(\fermionbackx,\fermionbacky){\oval(2200,2200)[r]}
\thicklines
\global\advance\fermionbackx by 1000
\drawarrow[\N\ATBASE](\fermionbackx,\fermionbacky)
\global\advance\fermionbackx by -2000
\drawarrow[\S\ATBASE](\fermionbackx,\fermionbacky)
\global\advance\fermionbackx by 1000
\global\advance\fermionbacky by 1000
\put(\fermionbackx,\fermionbacky){\circle*{500}}
\thinlines
\drawline\fermion[\N\REG](\fermionbackx,\fermionbacky)[1500]
\thicklines
\drawarrow[\LDIR\ATBASE](\pmidx,\pmidy)
\put(18000,3500){\makebox(0,0){$+$}}
\thinlines
\drawline\fermion[\E\REG](22000,3000)[4000]
\put(\fermionbackx,\fermionbacky){\circle*{500}}
\thicklines
\drawarrow[\LDIR\ATTIP](\pmidx,\pmidy)
\thinlines
\global\advance\fermionbacky by 100
\drawline\fermion[\E\REG](\fermionbackx,\fermionbacky)[\fermionlength]
\global\advance\fermionfronty by -200
\drawline\fermion[\E\REG](\fermionfrontx,\fermionfronty)[\fermionlength]
\global\advance\pmidy by 100
\thicklines
\drawarrow[\LDIR\ATBASE](\pmidx,\pmidy)
\thinlines
\global\advance\fermionbacky by 100
\put(\fermionbackx,\fermionbacky){\circle*{500}}
\drawline\fermion[\E\REG](\fermionbackx,\fermionbacky)[\fermionlength]
\thicklines
\drawarrow[\LDIR\ATBASE](\pmidx,\pmidy)
\thinlines
\drawline\fermion[\N\REG](\fermionfrontx,\fermionfronty)[2000]
\thicklines
\drawarrow[\LDIR\ATBASE](\pmidx,\pmidy)
\thinlines
\put(\fermionbackx,\fermionbacky){\circle*{500}}
\drawline\fermion[\W\REG](\fermionbackx,\fermionbacky)[4000]
\thicklines
\drawarrow[\E\ATBASE](\pmidx,\pmidy)
\thinlines
\put(\pmidx,\pmidy){\oval(3800,1800)[t]}
\put(\pmidx,\pmidy){\oval(4200,2200)[t]}
\thicklines
\global\advance\pmidy by 1000
\drawarrow[\W\ATBASE](\pmidx,\pmidy)
\thinlines
\put(\fermionbackx,\fermionbacky){\circle*{500}}
\drawline\fermion[\S\REG](\fermionbackx,\fermionbacky)[2000]
\thicklines
\drawarrow[\LDIR\ATBASE](\pmidx,\pmidy)
\thinlines
\drawline\fermion[\E\REG](22000,1100)[12000]
\global\advance\fermionfronty by -200
\drawline\fermion[\E\REG](\fermionfrontx,\fermionfronty)[\fermionlength]
\global\advance\pmidy by 100
\thicklines
\drawarrow[\LDIR\ATBASE](\pmidx,\pmidy)
\end{picture}}
\put(0,8000)
{\begin{picture}(40000,4000)
\put(1000,2000){\makebox(0,0){$+$}}
\thinlines
\drawline\fermion[\E\REG](2000,3000)[3000]
\put(\fermionbackx,\fermionbacky){\circle*{500}}
\thicklines
\drawarrow[\LDIR\ATTIP](\pmidx,\pmidy)
\thinlines
\global\advance\fermionbackx by 1500
\put(\fermionbackx,\fermionbacky){\oval(3000,2000)[t]}
\put(\fermionbackx,\fermionbacky){\oval(2800,1800)[b]}
\put(\fermionbackx,\fermionbacky){\oval(3200,2200)[b]}
\thicklines
\global\advance\fermionbacky by 1000
\drawarrow[\W\ATBASE](\fermionbackx,\fermionbacky)
\global\advance\fermionbacky by -2000
\drawarrow[\E\ATBASE](\fermionbackx,\fermionbacky)
\global\advance\fermionbackx by 1500
\global\advance\fermionbacky by 1000
\thinlines
\put(\fermionbackx,\fermionbacky){\circle*{500}}
\drawline\fermion[\E\REG](\fermionbackx,\fermionbacky)[3000]
\put(\fermionbackx,\fermionbacky){\circle*{500}}
\thicklines
\drawarrow[\LDIR\ATBASE](\pmidx,\pmidy)
\thinlines
\global\advance\fermionbacky by 100
\drawline\fermion[\E\REG](\fermionbackx,\fermionbacky)[\fermionlength]
\global\advance\fermionfronty by -200
\drawline\fermion[\E\REG](\fermionfrontx,\fermionfronty)[\fermionlength]
\global\advance\pmidy by 100
\thicklines
\drawarrow[\LDIR\ATBASE](\pmidx,\pmidy)
\global\advance\fermionfronty by 100
\thinlines
\drawline\fermion[\S\REG](\fermionfrontx,\fermionfronty)[2000]
\thicklines
\drawarrow[\N\ATBASE](\pmidx,\pmidy)
\thinlines
\put(\fermionbackx,\fermionbacky){\circle*{500}}
\drawline\fermion[\E\REG](\fermionbackx,\fermionbacky)[3000]
\thicklines
\drawarrow[\LDIR\ATBASE](\pmidx,\pmidy)
\thinlines
\global\advance\fermionfronty by 100
\drawline\fermion[\W\REG](\fermionfrontx,\fermionfronty)[9000]
\global\advance\fermionfronty by -200
\drawline\fermion[\W\REG](\fermionfrontx,\fermionfronty)[\fermionlength]
\thicklines
\global\advance\pmidy by 100
\drawarrow[\E\ATBASE](\pmidx,\pmidy)
\put(18000,2000){\makebox(0,0){$+$}}
\thinlines
\drawline\fermion[\E\REG](22000,3000)[3000]
\put(\fermionbackx,\fermionbacky){\circle*{500}}
\thicklines
\drawarrow[\LDIR\ATTIP](\pmidx,\pmidy)
\thinlines
\global\advance\fermionbacky by 100
\drawline\fermion[\E\REG](\fermionbackx,\fermionbacky)[9000]
\global\advance\fermionfronty by -200
\drawline\fermion[\E\REG](\fermionfrontx,\fermionfronty)[\fermionlength]
\global\advance\pmidy by 100
\thicklines
\drawarrow[\LDIR\ATBASE](\pmidx,\pmidy)
\global\advance\fermionfronty by 100
\thinlines
\drawline\fermion[\S\REG](\fermionfrontx,\fermionfronty)[2000]
\thicklines
\drawarrow[\N\ATBASE](\pmidx,\pmidy)
\put(\fermionbackx,\fermionbacky){\circle*{500}}
\thinlines
\global\advance\fermionbacky by 100
\drawline\fermion[\W\REG](\fermionbackx,\fermionbacky)[3000]
\global\advance\fermionfronty by -200
\drawline\fermion[\W\REG](\fermionfrontx,\fermionfronty)[\fermionlength]
\thicklines
\global\advance\pmidy by 100
\drawarrow[\E\ATBASE](\pmidx,\pmidy)
\global\advance\fermionfronty by 100
\thinlines
\drawline\fermion[\E\REG](\fermionfrontx,\fermionfronty)[\fermionlength]
\thicklines
\drawarrow[\LDIR\ATTIP](\pmidx,\pmidy)
\thinlines
\put(\fermionbackx,\fermionbacky){\circle*{500}}
\global\advance\fermionbackx by 1500
\put(\fermionbackx,\fermionbacky){\oval(3000,2000)[t]}
\put(\fermionbackx,\fermionbacky){\oval(2800,1800)[b]}
\put(\fermionbackx,\fermionbacky){\oval(3200,2200)[b]}
\thicklines
\global\advance\fermionbacky by 1000
\drawarrow[\W\ATBASE](\fermionbackx,\fermionbacky)
\global\advance\fermionbacky by -2000
\drawarrow[\E\ATBASE](\fermionbackx,\fermionbacky)
\global\advance\fermionbackx by 1500
\global\advance\fermionbacky by 1000
\thinlines
\put(\fermionbackx,\fermionbacky){\circle*{500}}
\drawline\fermion[\E\REG](\fermionbackx,\fermionbacky)[3000]
\thicklines
\drawarrow[\LDIR\ATBASE](\pmidx,\pmidy)
\end{picture}}
\put(0,2000)
{\begin{picture}(40000,4000)
\put(1000,2000){\makebox(0,0){$+$}}
\put(3000,1000){\makebox(0,0){$\dots$}}
\put(38000,2000){\makebox(0,0){$(a\ge3)$}}
\end{picture}}
\end{picture}
}
\vspace{-0.5cm}
\caption{\label{Dysonreihe}
\sf Feynman diagrams of the series (\ref{bary_prop}) for the baryon
propagator $G^{val}_B$}
\end{figure}

Up to this point, we have dealt with the generating functional
(\ref{gener_func}) for meson, diquark and baryon fields. Below we
derive an effective meson-baryon action generalizing the approach of
\cite{Rei90}. For this purpose we perform the diquark integration.
Since the diquark fields appear to all orders we introduce external
diquark sources $j,\tilde{j}$ and rewrite $Z \sim  \int \D \varphi \D 
\bar{\Psi}^\alpha \D \Psi^\alpha Z_{diq}[j,\tilde{j}]$
(\ref{gener_func}) in the form  
\be
\label{Z_diq}
 Z_{diq}[j,\tilde{j}] &\sim& \int \D \Delta \D \tilde{\Delta} \exp
 \left( \A^{(2)}[\Delta, \tilde{\Delta}] + \A_{int}[\Delta,
 \tilde{\Delta}] + i \left[ \tilde{\Delta} j + \tilde{j} \Delta
 \right] \right) \nonumber \\ 
 &\sim& \exp \left( \A_{int}[\frac{\delta}{i\delta\tilde{j}} ,
 \frac{\delta}{i\delta j} ] \right) \int \D \Delta \D \tilde{\Delta}
 \exp \left( \tilde{\Delta} D_{diq}^{-1} \Delta +
 i (\tilde{\Delta} j + \tilde{j} \Delta) \right) .
\ee
The integration over the diquark fields results in
\be
 Z_{diq}[j,\tilde{j}] &\sim& \exp \left(
 \A_{val}[\frac{\delta}{i\delta\tilde{j}} , 
 \frac{\delta}{i\delta j} ] \right) \exp \left(
 \tilde{j} D_{diq} j \right) .
\ee
Note, that the interaction part $\A_{int}=\A_{val}$ is completely
determined by the valence quark part of the baryon propagator. Using
the geometric series
\be
 \left( 1 + \G_0 \Delta \tilde{\G_0} \tilde{\Delta}
 \right)^{-1} &=& \left( 1 + \G_0 b_\gamma \Gamma_{diq}^\gamma
 \tilde{\G_0} \tilde{\Gamma}_{diq}^\delta b^*_\delta
 \right)^{-1} \nonumber \\
 &=& \sum_{a=0}^{\infty} \left(\G_0 b_\gamma \Gamma_{diq}^\gamma
 \tilde{\G_0} \tilde{\Gamma}_{diq}^\delta b^*_\delta \right)^a
\ee
to reexpress $G_{11}$ in eq.~(\ref{G_11}) we obtain the baryon
propagator  
\be
\label{bary_prop}
 (G^{val}_B)_{\alpha \beta} = \G_0 D_{\alpha \beta} &+& \G_0
 D_{\gamma\delta} \Gamma_{diq}^\gamma \tilde{\G_0}
 \tilde{\Gamma}_{diq}^\delta  D_{\alpha \beta} \G_0 \nonumber \\
 &+& \G_0 D_{\alpha\delta} \Gamma_{diq}^\gamma
 \tilde{\G_0} \tilde{\Gamma}_{diq}^\delta D_{\gamma\beta} \G_0 + \dots\ .
\ee
as an infinite series in quark $\G_0$ and diquark propagators
$D_{\alpha \beta}$. This series is drawn as a sum of Feynman diagrams in
Fig.~\ref{Dysonreihe}.  
For the calculation of bound states it is only necessary to consider
the quark exchange graphs in the series (\ref{bary_prop}). Neglecting
the self-interaction graphs is justified because we have an effective
interaction and therefore all self-interaction graphs can be absorbed
in the effective coupling constants. This simplification is nothing
else than the well-known ladder approximation which has the advantage
that the baryon propagator can summed up completely: 
\be
\label{Baryonprop}
 (G^{val}_B)_{\alpha \beta} &=& \G_0 D_{\alpha \beta} + \G_0
 D_{\alpha\delta} \Gamma_{diq}^\gamma \tilde{\G_0} \tilde{\Gamma}_{diq}^\delta
 D_{\gamma\beta} \G_0 \nonumber \\
 &&+ \G_0 D_{\alpha\delta} \Gamma_{diq}^\gamma \tilde{\G_0}
 \tilde{\Gamma}_{diq}^\delta D_{\gamma\eta} \Gamma_{diq}^\epsilon \tilde{\G_0}
 \tilde{\Gamma}_{diq}^\eta D_{\epsilon\beta} \G_0 + \dots \nonumber \\
 &=& (G^0_B)_{\alpha \beta} + (G^0_B)_{\alpha\delta}
 H^{\gamma\delta} (G^0_B)_{\gamma\beta} +
 (G^0_B)_{\alpha\delta} H^{\gamma\delta}
 (G^0_B)_{\gamma\eta} H^{\epsilon\eta} (G^0_B)_{\epsilon\beta} +
 \dots \nonumber \\
 &=& \left( 1- (G^0_B)_{\alpha\delta} H^{\gamma\delta} \right)^{-1}
 (G^0_B)_{\gamma\beta} .
\ee
Here we have introduced the free quark-diquark propagator
\be
 (G^0_B)_{\alpha\beta} =\G_0 D_{\alpha \beta}
\ee
and the quark exchange operator
\be
 H^{\gamma\delta} = \Gamma_{diq}^\gamma \tilde{\G_0}
 \tilde{\Gamma}_{diq}^\delta .
\ee
We are now able to integrate over the baryon source fields
$\Psi^\alpha , \bar{\Psi}^\alpha$. This leads to an effective baryon
action  
\be
 \A^{(2)}_{bary} = \bar{B}^\alpha (G^{val}_B)^{-1}_{\alpha \beta}
 B^\beta
\ee
for the baryon fields $\bar{B}^\alpha=P^\nu \bar{B}^\alpha_\nu$ und
$B^\alpha = B^\alpha_\nu P^\nu$. The inversion of the baryon propagator 
\be
 (G^{val}_B)^{-1}_{\alpha \beta} = (G^0_B)^{-1}_{\alpha\beta} -
 H_{\alpha\beta}
\ee
follows directly from eq.~(\ref{Baryonprop}). Using 
\be
 \frac{\delta\A^{(2)}_{bary}}{\delta\bar{B}^\alpha}=0
\ee
one finally arrives at the Faddeev equation in operator form:
\be
 (G^{val}_B)^{-1}_{\alpha \beta}  B^\beta = 0 .
\ee

In the next step we construct the kernel of the matrix
$(G^{val}_B)^{-1}_{\alpha \beta}$. For this purpose we expand the
baryon field 
\be
\label{ProdAnsatz}
 B(x,y) = \int d\omega dE 
 \sum_{\alpha}\sum_{\nu}^{(\epsilon_\nu>0)} 
 a_{\alpha\nu}(E,\omega) B_{\alpha\nu}^{0}(\vec{x},x_4;\vec{y},y_4)
\ee
in eigenfunctions 
\be
 B_{\alpha\nu}^{0}(\vec{x},x_4;\vec{y},y_4) = \frac{1}{(2\pi)^2}
 \Delta_\alpha(\vec{x}) e^{-i \omega x_4} \Psi_\nu(\vec{y}) e^{-iEy_4}
\ee
of the free quark-diquark propagator
\be
 (G^0_B)^{-1}_{\alpha \beta} B_{\beta\nu}^{0}(\vec{x},x_4;\vec{y},y_4)
 = - \beta \left( iE-\epsilon_\nu \right) \left( \omega^2 + \omega_\beta^2
 \right) B_{\beta\nu}^{0}(\vec{x},x_4;\vec{y},y_4).
\ee
The projection operator $P^\beta$ (\ref{baryon}) on the third quark
projects out the positive energy eigenstates $(\epsilon_\nu>0)$. 
Transforming the energy variables $E,\omega$ to the total energy
$\Omega=E+\omega$ and the relative energy $\tilde{\Omega} =
\frac{1}{2}(\omega-E)$  we obtain the Faddeev equation in terms of the
coefficients $a_{\alpha\nu}(i\Omega,i\tilde{\Omega})$:  
\be
\label{Fad-gl}
 &&\left[ \frac{1}{2}i\Omega -i \tilde{\Omega}  -\epsilon_\nu
 \right] \left[ \left( \frac{1}{2}\Omega + \tilde{\Omega}\right)^2 +
 \omega_\alpha^2 \right] a_{\alpha\nu} (i\Omega,i\tilde{\Omega}) \\
 \nonumber 
 && \qquad \qquad \qquad \qquad + \frac{1}{4} \int
 d\tilde{\Omega}^\prime \sum_{\beta,\mu,\kappa} \frac{\langle \nu | 
 \Gamma^\beta_{diq} | \kappa \rangle \langle \kappa |
 \Gamma^\alpha_{diq} | \mu \rangle}{i \left(\tilde{\Omega}  +
 \tilde{\Omega}^\prime\right)  + \epsilon_\kappa}  a_{\beta\mu}
 (i\Omega,i\tilde{\Omega}^\prime)= 0. 
\ee
The summation runs over the one-particle quark eigenstates $| \mu
\rangle$ and $| \kappa \rangle$ of $h_\Theta$. To reduce the numerical
effort we restrict the diquark states to the energetically lowest
state $\Gamma^\alpha_{diq}=\Gamma_{diq}$\footnote{In the following we 
suppress all diquark indices, \eg
$a_{\alpha,\mu}(i\Omega,i\tilde{\Omega}) = a_\mu 
(i\Omega,i\tilde{\Omega})$.}. In fact, this is the only bound 
state; all other states lie above the two quark threshold and,
if at all, influence the quark-diquark bound state only weakly.  

It is important to equip the absolute and relative energies, $\Omega$
and $\tilde{\Omega}$, respectively with a physical meaning. These
quantities do not yet have apparent interpretations since due to the
presence of the static and localized soliton Lorentz covariance is
lost. Hence we are still lacking an energy scale. In order to define a
physically relevant energy scale we consider the limit of vanishing 
diquark-quark coupling ($\Gamma_{diq}\rightarrow0$). In this limit the
different quark states decouple and the Faddeev equation reduces to a
purely algebraic equation. For the valence quark part we have the
three solutions $i\Omega=2(i\tilde{\Omega}+\epsilon_{val})$ and
$i\Omega=4(-i\tilde{\Omega}\pm\omega_{diq})$. When the energy transfer
vanishes, \ie $\tilde{\Omega}=0$, we should obtain the physically
reasonable solution $i\Omega=\omega_{diq}+\epsilon_{val}$ for the
energy of a free baryon. This result can only be realized if we
enforce the variable substitution $\tilde{\Omega} \rightarrow
\tilde{\Omega} + \frac{i}{2}(\epsilon_{val} - \omega_{diq})$. Thus,
two of the three solutions join the physical solution. The third one
yielding a negative solution is rejected because we have projected to
positive energy states. Since the limit of vanishing diquark-quark
coupling should continuously emerge from the complete solution of the
Faddeev equation it is necessary to perform the variable
substitution\footnote{A similar variable substitution was implemented 
by Ishii et.~al \cite{Ish95}.} $\tilde{\Omega} \rightarrow
\tilde{\Omega} + \frac{i}{2}(\epsilon_{val} - \omega_{diq})$ also in
the Faddeev equation (\ref{Fad-gl}). 

For the numerical solution of this integral equation\footnote{A
similar approach was employed for the solution of the Bethe-Salpeter
equation for the diquarks\cite{Zue95a}.} we discretize the integration
variable $\tilde{\Omega}_j = j \Delta \tilde{\Omega}; j=0,\dots,N$,
yielding the kernel as an $N \times N$ matrix. We then generalize the
matrix equation to the eigenvalue equation  
\be
\label{Fad-diskr}
 & &\left[ \frac{1}{2}\left( i\Omega + \epsilon_{val} - 2 \epsilon_\nu -
 \omega_{diq} \right)  -i \tilde{\Omega}_j  \right] \left[ \left(
 \tilde{\Omega}_j +  \frac{i}{2} \left\{ \epsilon_{val} - \omega_{diq}
 - i \Omega \right\} \right)^2 + \omega_{diq}^2 \right] a_{\nu}
 (i\Omega,i\tilde{\Omega}_j) \nonumber \\
 && \qquad + \frac{1}{4} \sum_{l,\kappa}
 \Delta \tilde{\Omega}  \frac{\langle \nu |
 \Gamma_{diq} | \kappa \rangle \langle \kappa |
 \Gamma_{diq} | \mu \rangle}{i \left(\tilde{\Omega}_j +
 \tilde{\Omega}_l \right) + \omega_{diq} + \epsilon_\kappa -
 \epsilon_{val}}  a_{\mu}(i\Omega,i\tilde{\Omega_j}) =
 \lambda_j(i\Omega) a_{\nu} (i\Omega,i\tilde{\Omega}_j) .
\ee
A solution of the Faddeev equation is given for a vanishing eigenvalue
$\lambda_j(i\Omega_B)=0$ with the energy $\Omega_B$. Within this
formulation there exists only one positive energy solution with
$\Omega_B < \epsilon_{val} + \omega_{diq}$. Possible negative
solutions will again be rejected. If we replace $\Gamma_{diq}$ in
(\ref{Fad-diskr}) with $\chi\Gamma_{diq}$ we can explore the
dependence of the solution on the coupling between quark and
diquark. In this manner we observe that $\Omega_B$ approaches the free
solution $\epsilon_{val} + \omega_{diq}$ for decreasing $\chi$. This
justifies the above described procedure.  

\section{Numerical results and discussion} 
\label{Numeric}

\begin{figure}
\centerline{
\epsfig{figure=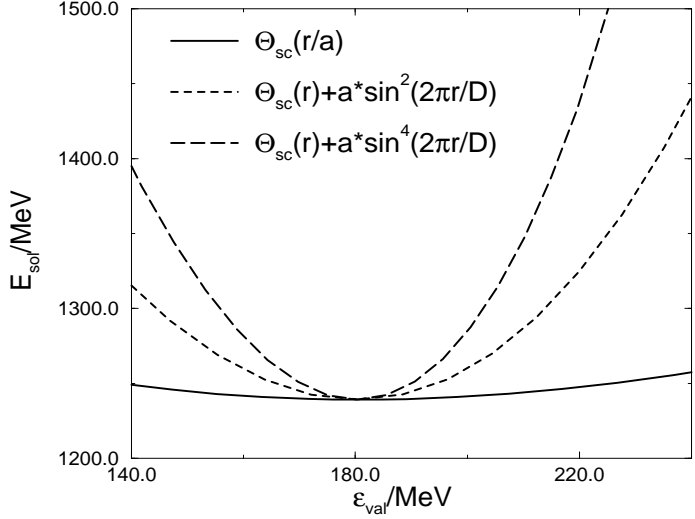,height=8.0cm,width=10.0cm}
}
\caption{\label{scaling}
\sf Dependence of the soliton energy from the valence quark energy for
three different parameterizations of the soliton profile. Results are
for $m=450$MeV.}
\end{figure}

In the first step we calculate\footnote{A complete description of the
numerical treatment is given in \cite{Alk94a}.} the chiral angle
$\Theta_{sc}(r)$ self-consistently in a box with radius $D$ by
extremizing the energy functional (\ref{ESol}). In order to later
estimate the influence of the quark-quark and the diquark-quark bound
states on the soliton we introduce a dimensionless parameter $a$,
which measures the extension of the soliton profile, via $\Theta(r) =
\Theta_{sc}(r/a)$. This choice is motivated by the softness of the
scaling mode, \ie a weak dependence of the total soliton energy on the
parameter $a$ in comparison to other variables. The energy of the
valence quark, on the other hand, depends sensitively on $a$. As shown
in Fig.~\ref{scaling} the soliton energy for the scaling mode is
almost constant in the displayed range. For comparison the
dependencies for two alternative parameterizations are presented. Both
profiles are chosen such that they not only fulfill the usual boundary
conditions but also the derivatives at the boundaries are unchanged.    

\begin{table}
\begin{tabular}{cccc} 
$a$ & $\omega_{diq}$/MeV & $2\epsilon_{val}$ & $B_{qq}$/MeV \\ \hline
0.0 & 759            & 934          & 175 \\ 
0.2 & 761            & 934          & 173 \\ 
0.4 & 758            & 929          & 171 \\ 
0.6 & 672            & 808          & 136 \\ 
0.8 & 465            & 560          &  95 \\ 
1.0 & 295            & 360          &  65 \\ 
1.2 & 161            & 208          &  47 \\ 
\end{tabular}
\bigskip
\caption{\label{diq_a}
\sf The energy $\omega_{diq}$ of the diquark in the soliton, the energy
$2\epsilon_{val}$ of two uncorrelated quarks and the binding energy of
the diquark $B_{qq}=2\epsilon_{val}-\omega_{diq}$ in dependence on the
scaling parameter $a$ for constituent quark mass $m=450$MeV and $g_2=g_1$.}
\end{table}

In the next step we solve the Bethe-Salpeter equation (\ref{BS}) for
the diquark fields\footnote{A test of the numerical treatment in the
finite box is given in appendix A.}. In Tab.~\ref{diq_a} we display
the solution of the Bethe-Salpeter equation for different soliton
profiles. The important result is that the diquarks are bound
($\omega_{diq} < 2 \epsilon_{val}$) for all considered
profiles. Furthermore, the binding energy $B_{qq} = 2 \epsilon_{val} -
\omega_{diq}$ decreases with growing soliton size. For a small soliton
extension ($a<0.4$) the diquark energy is almost independent of the
soliton. Note, that the limit $a \rightarrow 0$ corresponds to a
diquark in an empty box. On the contrary, for larger soliton extension
($a>0.5$) we observe a strong dependence on the scaling factor
$a$. This behavior of the diquark energy is inherited from the one of
the valence quark energy. This indicates that the properties of the
diquark are essentially determined by the valence quarks. 

\begin{table}
\begin{tabular}{cccc} 
$m$/MeV & $\omega_{diq}$/MeV & $2\epsilon_{val}$ & $B_{qq}$/MeV \\ \hline
350 & 455            & 495          &  40 \\ 
400 & 368            & 421          &  53 \\ 
450 & 295            & 360          &  65 \\ 
500 & 225            & 303          &  78 \\ 
600 &  80            & 189	    & 109 \\ 
\end{tabular}
\bigskip
\caption{\label{diq_m}
\sf The same as Tab.~\protect\ref{diq_a} in dependence on the constituent
quark mass $m$ for the self-consistent soliton and $g_2=g_1$.}
\end{table}

A further interesting point is the increasing diquark binding energy
with growing constituent quark mass when the self-consistent soliton
is employed as can be seen from Tab.~\ref{diq_m}. This feature is
particularly astonishing because the valence quark energy decreases
simultaneously. 

\begin{figure}
\centerline{
\epsfig{figure=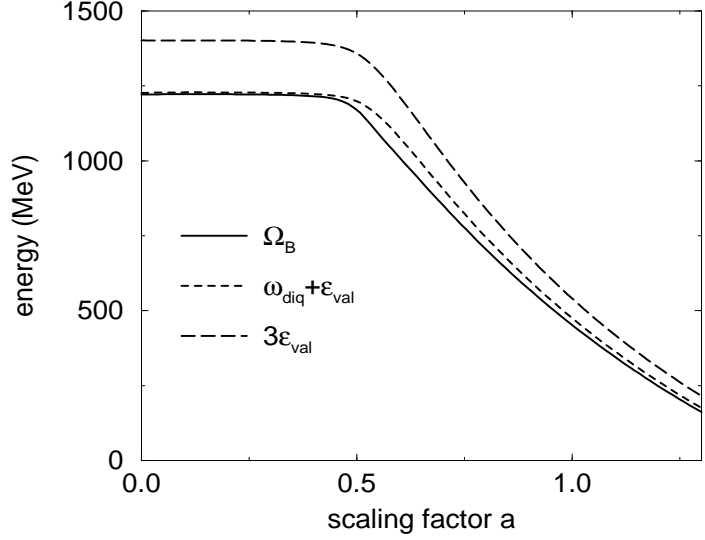,height=8.0cm,width=10.0cm}
}
\caption{\label{fad1}
\sf Solution of the Faddeev equation (solid line), the energy of a
free diquark-quark pair (short dashed line) and the energy of three
uncorrelated valence quarks (long dash line) in dependence on the
scaling factor $a$ for a constituent quark mass $m=450$MeV and $g_2=g_1$.}
\end{figure}

These results for the diquark serve as input to the solution of the
Faddeev equation (\ref{Fad-diskr}). In Fig.~\ref{fad1} we display the
resulting energy $\Omega_B$ as a function of the scaling variable $a$
for a constituent quark mass $m=450$MeV. For comparison, the energy of
three uncorrelated valence quarks and the energy of a free
diquark-quark pair are also shown. Again, in the entire region the
diquark and quark are bound, although only weakly. In
Tab.~\ref{Bindenerg} we compare the quark-quark ($B_{qq} =
2\epsilon_{val} - \omega_{diq}$) and diquark-quark ($B_{qd} =
\epsilon_{val} + \omega_{diq} - \Omega_B$) binding energies to the
total binding energy 
\be
 B_{tot}=3\epsilon_{val}-\Omega_B = B_{qq} + B_{qd}
\ee
for a diquark coupling constant $g_2=g_1$. We notice that the binding
energy is carried to a large extent by the quark-quark correlations.
While these have their maximal value for a vanishing soliton ($a \to
0$) the residual diquark-quark interaction develops a maximum for
moderate soliton extension. More distinct is the situation
(Tab.~\ref{g2=2g1}) for a diquark coupling constant twice as large,
\ie $g_2 = 2 g_1$. In particular, the total binding energy is strongly
dominated by the diquark binding and is maximal for a vanishing
soliton. 

\begin{table}
\begin{tabular}{ccccc} 
$a$ & $\Omega_B$/MeV & $B_{qq}$/MeV & $B_{qd}$/MeV & $B_{tot}$/MeV \\ \hline
0.0 & 1221           & 175          & 6            & 181       \\ 
0.2 & 1222           & 173          & 6            & 179       \\ 
0.4 & 1215           & 171          & 8            & 179       \\ 
0.6 & 1010           & 136          & 66           & 202       \\ 
0.8 &  705           & 95           & 40           & 135       \\ 
1.0 &  452           & 65           & 24           & 89        \\ 
1.2 &  249           & 47           & 15           & 62        \\ 
\end{tabular}
\bigskip
\caption{\label{Bindenerg}
\sf The solution $\Omega_B$ of the Faddeev equation as well as the
quark-quark ($B_{qq}$) and diquark-quark binding energy $B_{qd}$ 
and the total binding energy $B_{tot}$ of the three valence quarks 
for a constituent quark mass $m=450$MeV and $g_2=g_1$.}
\end{table}

\begin{table}
\begin{tabular}{ccccc}
$a$ & $\Omega_B$/MeV & $B_{qq}$/MeV & $B_{qd}$/MeV &
$B_{tot}$/MeV \\ \hline
0.0 & 952            & 433          &  16          & 449       \\ 
0.2 & 957            & 428          &  17          & 445       \\ 
0.4 & 949            & 401          &  43          & 444       \\ 
0.6 & 787            & 315          & 110          & 425       \\ 
0.8 & 547            & 169          & 124          & 293       \\ 
1.0 & 374            & 108          &  59          & 167       \\ 
1.2 &  249           &  75          &  36          & 111       \\ 
\end{tabular}
\bigskip
\caption{\label{g2=2g1}
\sf The same as Tab.~\protect\ref{Bindenerg} for the coupling constant
$g_2=2g_1$.} 
\end{table}

In the past some authors have used instantaneous or static
approximations to solve the integral equation
\cite{Buc92,Hua94,Mey94,Han95,Kei96}. The instantaneous approximation  
corresponds to a quark exchange with vanishing relative energy
$\tilde{\Omega} = 0$ in eq.~(\ref{Fad-diskr}). Then the different
quark channels decouple and the Faddeev equation simplifies to a
purely algebraic equation:  
\be
\label{inst}
  \left[ i \Omega - \epsilon_{val} - \omega_{diq}
  \right] \left[  \left( \epsilon_{val} - \omega_{diq}  - i \Omega
  \right)^2 - 4 \omega_{diq}^2 \right] - 2 \sum_{\kappa}\frac{\langle val |
 \Gamma_{diq} | \kappa \rangle \langle \kappa |
 \Gamma_{diq} | val \rangle} {\omega_{diq} +\epsilon_\kappa -
 \epsilon_{val}} = 0 .
\ee
In the static approximation in addition the sum over the exchanged
quark is restricted to the valence quark:
\be
\label{static}
  \left[ i\Omega - \epsilon_{val} - \omega_{diq}
  \right] \left[ \left(   \epsilon_{val} - \omega_{diq}
 - i\Omega \right)^2 - 4 \omega_{diq}^2 \right] - 2\frac{\langle val |
 \Gamma_{diq} | val \rangle \langle val |
 \Gamma_{diq} | val \rangle} {\omega_{diq} } = 0 .
\ee
The presence of the energy denominator indicates that these
approximations should be well suited for large quark and diquark
masses. In fact, Tab.~\ref{Statvgl} confirms this assessment. For a
vanishing soliton the error is one per cent. When the soliton
extension grows the valence quark as well as the diquark energy become
smaller and therefore the error rises to about 20$\%$. This emphasizes
the importance of a complete solution of the Faddeev equation for a
solitonic background. In addition, this table shows only a small
difference between both approximations. In comparison to the complete
solution of the Faddeev equation both treatments possess an enhanced
binding energy. Hence we conclude that retardation effects in the full
calculation have a repulsive character. Both approximations also
elucidate the importance of the above described variable substitution,
because the free solution $i\Omega=\omega_{diq}+\epsilon_{val}$ is
obvious for vanishing diquark-quark coupling ($\Gamma_{diq}
\rightarrow 0$). If we improperly had carried out the static
approximation in equation (\ref{Fad-gl}) we would have had two
unphysical solutions $i\Omega=2\epsilon_{val}$ and $i\Omega =
\pm\omega_{diq}$.   

\begin{table}
\begin{tabular}{cccc} 
$a$ & $\Omega_B$/MeV & $\Omega_B^{(stat)}$/MeV &
$\Omega_B^{(inst)}$/MeV \\ \hline
0.0 & 1221           & 1202         & 1201         \\ 
0.2 & 1222           & 1204         & 1202         \\ 
0.4 & 1215           & 1194         & 1189         \\ 
0.6 & 1010           & 935          & 924          \\ 
0.8 &  705           & 630          & 628          \\ 
1.0 &  452           & 394          & 393          \\ 
1.2 &  249           & 203          & 202          \\ 
\end{tabular}
\bigskip
\caption{\label{Statvgl}
\sf The Solution $\Omega_B$ of the Faddeev equation
({\protect\ref{Fad-diskr}}) as well as the solution of the static
approximation $\Omega_B^{(stat)}$ (\protect{\ref{static}}) and the
instantaneous approximation $\Omega_B^{(inst)}$
({\protect\ref{inst}}) in dependence of the scaling factor $a$.} 
\end{table}

\section{A hybrid model for baryons}
\label{Hybrid}

\begin{figure}
\centerline{
\epsfig{figure=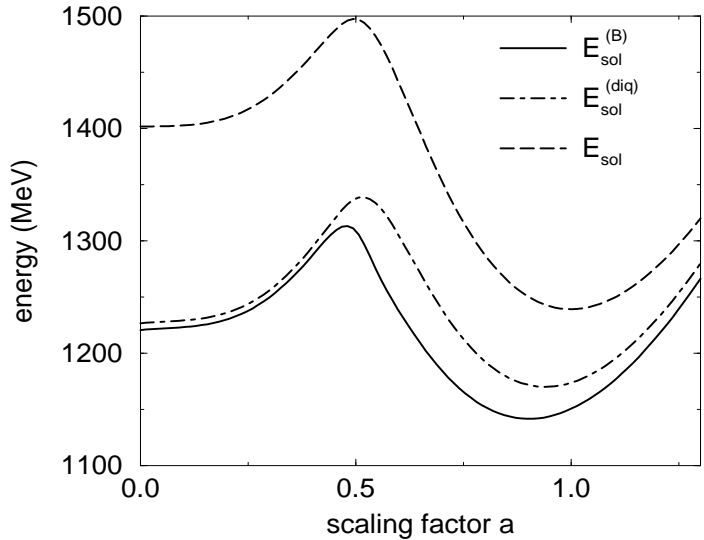,height=8.0cm,width=10.0cm}
}
\caption{\label{bar1} 
\sf The energy of the soliton with correlated valence quarks (solid
line), the energy of the additive diquark-quark--soliton model
(dash-dotted line) and the energy of the soliton with uncorrelated
valence quarks (long dashed line) in dependence of the scaling factor
$a$ for a constituent quark mass $450$MeV.} 
\end{figure}

Given the results of the last section we now consider a soliton
configuration with correlated valence quarks. For that purpose we
replace the contribution $3\epsilon_{val}$ of the three uncorrelated
valence quarks in the energy functional (\ref{ESol}) by the solution
of the Faddeev equation. The total energy of this configuration is
given by  
\be
\label{EFad}
 E_{sol}^{(B)} = E_{sea}+\Omega_{B} = E_{sol}- B_{tot} <
 E_{sol} ( = E_{sea}+ 3 \epsilon_{val})
\ee
and is shown in Fig.~\ref{bar1} as a function of the scaling parameter
$a$ for the constituent quark mass of $450$MeV. For comparison, the
energy of the pure soliton (\ref{ESol}) $ E_{sol} = E_{sea} + 3
\epsilon_{val}$ and the energy of the additive diquark-quark soliton
model $E_{sol}^{(diq)} = E_{sea}+\omega_{diq}+\epsilon_{val}$ defined
in \cite{Zue95a} are also shown. As compared to the additive
diquark-quark soliton picture the soliton with correlated valence
quarks has a lower energy in the entire range shown. The minimum of
this curve defines the hybrid model. Two other points in this figure
also have a special meaning: First, the point $a=0$ on the same curve
marks the model where a baryon is described as a bound state of a
quark and a diquark with the meson fields fixed at their vacuum
expectation values \cite{Mey94,Ish95}. Note, that this configuration
only corresponds to a local minimum of the energy. The other important
point is the minimum of $E_{sol}$ at $a=1$ which is nothing else than
the self-consistent soliton solution without any valence quark
correlations \cite{Alk96a}. Also the energy of this configuration is
larger than the energy of the hybrid model.  

\begin{table}
\begin{tabular}{lccccc}
$m$ (MeV)           & 350  &  400  &  450  &  500  &  600  \\ \hline
$E_{sol}^B$ (MeV)   & 1057 & 1140  & 1142  & 1108  &  1011 \\
$E_{sea}$ (MeV)     &   0  &    0  &  570  &  569  &  606  \\
$\Omega_B$ (MeV)    & 1057 & 1140  &  572  &  539  &  405  \\
$a$                 &   0  &    0  &  0.90 &  0.85 &  0.80 \\
$\omega_{diq}$ (MeV)& 688  &  725  &  375  &  356  &  283  \\ 
$\sqrt{\langle r_{I=0}^2 \rangle}$ 
                    & ---  &  ---  &  0.56 & 0.53  &  0.47 \\
$g_A$               & 0.32 & 0.32  &  0.70 & 0.68  &  0.64 \\
\end{tabular}
\bigskip
\caption{\label{Fad-Sol}
\sf The total energy of the lowest energy configuration $E_{sol}^B$ as
well as the contribution from the Dirac sea $E_{sea}$ and from the
correlated valence quarks $\Omega_B$ for $g_2=g_1$. Furthermore
displayed is the scaling parameter $a$, the effective energy
$\omega_{diq}$ of the diquark, the isoscalar radius $\langle r_{I=0}^2
\rangle^{1/2}$ and the axial coupling constant $g_A$.}
\end{table}

In Table \ref{Fad-Sol} we display results for the lowest energy
configuration for different constituent quark masses. This table shows
that for a large enough constituent quark mass the hybrid soliton in
fact is the minimal energy configuration. The lower limit of stability
for the hybrid soliton is around $m\approx430$MeV. Below this value
the energy of the baryon $E_{sol}^{(B)}=\Omega_B$ is carried
completely by the valence quarks. Herein, the diquark possesses an
effective energy of about $700$MeV. In comparison, for $m \approx
450$MeV the total energy of the self-consistent soliton configuration
is composed half by the valence quarks and half by the polarized
sea. The effective diquark energy drops down to the range
$280-380$MeV.    

\begin{table}
\begin{tabular}{dddddd} 
$a$ & $g_A^{val}$ & $g_A^{Fad}$ & $g_A^{diq}$ & $g_A^{sea}$ & $g_A$ \\ \hline
0.0 & 0.325 & 0.325 & 0.    &  0.    & 0.325 \\
0.2 & 0.325 & 0.325 & 0.001 & -0.001 & 0.325 \\
0.4 & 0.324 & 0.323 & 0.017 & -0.012 & 0.328 \\
0.6 & 0.273 & 0.253 & 0.478 & -0.033 & 0.698 \\
0.8 & 0.246 & 0.239 & 0.485 & -0.032 & 0.692 \\
0.9 & 0.240 & 0.236 & 0.483 & -0.018 & 0.701 \\
1.0 & 0.237 & 0.234 & 0.483 &  0.005 & 0.722 \\
1.1 & 0.235 & 0.233 & 0.485 &  0.035 & 0.753 \\
1.2 & 0.234 & 0.232 & 0.489 &  0.074 & 0.795 \\
\end{tabular}
\bigskip
\caption{\label{g_A}
\sf Contributions to the axial coupling constant $g_A$ in dependence
of the scaling factor $a$. For comparison also the contribution
$g_A^{val}$ of an uncorrelated valence quark is displayed.} 
\end{table}

In Table \ref{Fad-Sol} we also present results for the isoscalar
radius $\langle r_{I=0}^2 \rangle^{1/2}$ and the axial coupling
constant $g_A$. These calculations\footnote{In the appendix C we give
                    a full description of the calculation of 
		    static observables in the hybrid model.} 
need some comments: The isoscalar radius is calculated as the spatial
expectation value  
\be
 \langle r_{I=0}^2 \rangle = \int d^3\mbox {\boldmath $r$} r^2
 \rho(\mbox {\boldmath $r$}) = 4 \pi \int_0^D dr r^4 \rho(r).
\ee
This expression is only well-defined for localized baryon densities. For
non-localized baryon densities the radius depends on the box size
$D$ which is the case when the soliton is absent\footnote{Because of
this volume divergence we state nothing in this case.}. For the
constituent quark mass $m=450$MeV the isoscalar radius in the hybrid
model ($0.56$fm) and the pure soliton model ($0.54$fm) are almost
identical. We conclude from this slight increase that the sea is less
attractive in the hybrid model despite of the stronger binding of the
valence quarks. Note that both predictions are close to the
experimental value $\langle r_{I=0}^2 \rangle^{1/2}= 0.62$fm. 

More interesting are the results for the axial coupling constant  
\be
 g_A = g_A^{Fad} + g_A^{diq} + g_A^{sea}
\ee
consisting of contributions from the sea part $g_A^{sea}$, the diquark
$g_A^{diq}$ and the residual bound valence quark $g_A^{Fad}$ (\cf
appendix C). These contributions are collected in Tab.~\ref{g_A} as a
function of the scaling factor $a$. For comparison we also display the
contribution $g_A^{val}$ of an uncorrelated valence quark. First, we
observe only a small difference between a free and a correlated
valence quark. As a function of the scaling factor we find two
distinct regions: For small $a$ (no soliton) we have $g^{Fad}_A
\approx 0.33$, for larger $a$ (soliton region) the value decreases to
$g^{Fad}_A\approx0.24$. On the other hand, the diquark contribution
shows the opposite behavior. It vanishes in the case of a trivial
background but it increases to more than twice the value of a single
valence quark in the presence of the soliton. Finally, in both regions
only small corrections occur, which originate from the sea
contribution. The vanishing diquark contribution for $a=0$ also
explains the small values for $m=350$MeV and $m=400$MeV in
Tab.~\ref{Fad-Sol}. In addition, these results indicate a missing
component to the axial coupling constant: axial diquarks should also
be included before comparing $g_A$ with the experimental value
$g_A=1.26$ \cite{Par94}. 

\begin{figure}
\centerline{
\epsfig{figure=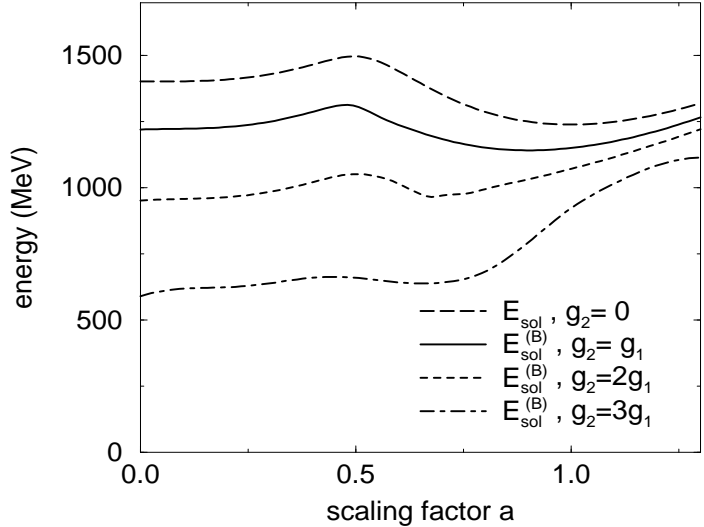,height=8.0cm,width=10.0cm}
}
\caption{\label{bar123} 
\sf The energy of the soliton with correlated valence quarks for
different diquark coupling constants for a constituent quark mass
$450$MeV.} 
\end{figure}

When we increase the diquark coupling $g_2$ we observe that stronger
diquark correlations counteract the formation of a soliton. As can be
seen from Fig.~\ref{bar123} the meson configuration, which minimizes
$E_{sol}^{(B)}$, is shrinking with increasing $g_2$. Starting at a 
critical coupling a path is open to a vanishing soliton. The displayed
curves correspond to four special cases. The pure soliton is given for
$g_2=0$, whereas the second one ($g_2=g_1$) describes the
Fierz-symmetric case. In the case $g_2=2g_1$ the energy
$E_{sol}^{(B)}$ is identical to the physical nucleon mass \cite{Buc92}
in the range where no soliton is present ($a<0.4$). Finally, in the
last case $g_2=3g_1$ we have an additional symmetry - the
Pauli-G\"ursey symmetry \cite{Pau57,Gue58,Vog91} - giving the diquarks
the same mass as the pions. However, this symmetry cannot be realized
in nature since otherwise the baryon masses should be much smaller. 

\section{Conclusions and outlook}
\label{Conclusions}

In this paper we have constructed a hybrid model for the description
of baryons within the framework of the NJL model. Using functional
integral techniques we have converted the pure quark model into an
effective theory of mesons, diquarks and baryons. As a first step we
have solved the Bethe-Salpeter equation for an $S$--wave scalar
diquark in the background of a soliton configuration. We have found
that this diquark is kinematically stable against the decay into the
two lightest quarks even in case the masses of the latter are strongly 
reduced by their interaction with the chiral soliton. In the next step
we have integrated out the diquark fields, thereby inducing an
interaction between quarks and the bound diquark through quark
exchange. Using the ladder approximation for this exchange interaction
a Faddeev type of equation has been derived, which includes the
solitonic background field. This equation yields a three--quark bound
state in the solitonic background. Using this result the hybrid has
been constructed combining the three--quark bound state with the
polarized sea. The energy of this hybrid is (for constituent quark
masses $m>430$MeV) smaller than either the energy of the soliton with
three uncorrelated quarks or the energy of the three--quark bound
state with no soliton present. Hence the formation of such a hybrid is
dynamically preferred as compared to either picture alone. In
addition, the soliton size of the hybrid is much smaller than the size
of the soliton with uncorrelated quarks, and the Faddeev amplitude,
\ie the baryon wave function in terms of quarks, significantly
deviates from the case without soliton. These results show that
baryons are presumably very much like hybrids containing both,
solitonic meson clouds and three--quark correlations, and that the
implementation of both features is crucial for a proper description of
baryons. Unfortunately, the question whether a baryon is dominated by
the valence quark structure or the soliton is still an open question. 
E.~g.~if one includes for the soliton also (axial-) vector mesons
the baryon current is completely carried by the polarized Dirac sea
\cite{Alk92,Zue95}. To clarify the situation one should in addition
include at the mean field level static diquark fields. The solution of
such an improved soliton calculation would show whether there is still
room for valence quark degrees of freedom beside vector meson degrees
of freedom or whether they would exclude each other. 
  
Within the hybrid model we have calculated the isoscalar radius
$\langle r_{I=0}^2 \rangle^{1/2}$ as well as the axial coupling
constant $g_A$. While the isoscalar radius shows reasonable values in  
comparison with experiment the inclusion of axial diquarks seems to be
important for the axial coupling constant. In order to calculate other
observables one has to perform the last step in the construction of
the hybrid model for baryons: one has to project the soliton solution
to good spin and flavor states.  

In this work we have considered the NJL model which is well suited for
low-energy properties of hadrons. Nevertheless, for energies higher
than twice the constituent quark mass, threshold effects associated
with the unphysical decay into free quarks disturb the calculation. 
Therefore, a generalization to bilocal chiral models seems necessary
in order to effectively incorporate quark confinement. Within these
models there can also exist confined diquarks \cite{Ben96a} on the one
hand. On the other hand, soliton calculations have shown that inside a
soliton propagating quarks \cite{Fra91,Tan91,Fra92} also can
exist. Thus, the hybrid model introduced in this work presents a well
suited starting point for further generalizations which appear
necessary to deepen our understanding of the structure of baryons. 
 
\acknowledgments 
{The authors are grateful to L.\ Gamberg for carefully reading the
manuscript. This work was partly supported by COSY under contract
41315266 and by Deutsche Forschungsgemeinschaft (DFG) under contract
Re 856/2-2.} 

\appendix

\section{Test of the numerical treatment in the finite box}

In order to test the numerical treatment in the finite
box we compare the solution of (\ref{BS}) for a vanishing chiral angle
$\Theta(r)=0$ with the solution in the continuum. The mass of the
diquark is obtained from the Bethe-Salpeter equation \cite{Wei93a}  
\be
\label{BS_Kont}
 \left[ -p^2 A(p^2) + 2 m^2 B - \frac{m_\pi^2 f_\pi^2}{6 m_0 m
 g_2/g_1} \right]_{p^2=-m_{diq}^2} = 0
\ee
with
\be
 A(p^2) &=& \frac{1}{16 \pi^2} \int_0^1 d\alpha
 \Gamma\left(0,\frac{m^2 + \alpha (1-\alpha) p^2}{\Lambda^2}\right),
 \nonumber \\
 B &=& \frac{1}{16 \pi^2} \Gamma\left(-1,\frac{m^2}{\Lambda^2}\right).
\ee

\begin{table}[t]
\begin{tabular}{ccccc} 
$g_2/g_1$ & $m_{diq}$/MeV & $\omega_{diq}$/MeV & 
$\omega_{diq}^{(1)}$/MeV & $\omega_{diq}^{(2)}$/MeV \\ \hline
0.0   & 900         & 934      & 934    & 909  \\ 
0.5   & 884         & 904      & 919    & 893  \\ 
1.0   & 736         & 759      & 778    & 747  \\ 
1.5   & 595         & 625      & 646    & 608  \\ 
2.0   & 463         & 502      & 527    & 480  \\ 
2.5   & 326         & 380      & 412    & 350  \\ 
3.0   & 140         & 241      & 288    & 188  \\ 
\end{tabular}
\bigskip
\caption{\label{Diquarkmasse}
\sf The mass $m_{diq}$ of a diquark in the continuum, the energy
$\omega_{diq}$ of a diquark as the solution of the Bethe-Salpeter
equation in a spherical box, $\omega_{diq}^{(1)}$ and
$\omega_{diq}^{(2)}$  from the approximation of
eq.~(\protect\ref{omega1}) and eq.~(\protect\ref{omega2}) respectively
in dependence of $g_2/g_1$.}
\end{table}

When we compare the solutions of the Bethe-Salpeter equations in the
continuum (\ref{BS_Kont}) and in the spherical box (\ref{BS}) we
observe a relatively large difference (\cf the 2$^{nd}$ and 3$^{rd}$
column in Tab.~\ref{Diquarkmasse}). However, we have to take into
account that in a finite box the quarks always possess finite momenta
and therefore an increased energy $\epsilon_{\mu} = \pm \sqrt{m^2+(\mu 
\pi/D)^2}$. Now, we can consider two limits: First, in the limit of
vanishing diquark coupling ($g_2\to 0$) a diquark consists of two free
quarks ($\omega_{diq}= 2 \epsilon_{val}$). From this we obtain the 
approximation  
\be
\label{omega1}
 \omega_{diq}^{(1)} &=& \sqrt{ m^2_{diq} + 4 (\pi/D)^2}
\ee
for the energy of a diquark in a spherical box. Second, we consider
the limit of a large quark-quark coupling. In this case, the diquark
is a highly correlated state which can be viewed as an independent
particle, hence 
\be
\label{omega2}
 \omega_{diq}^{(2)} &=& \sqrt{ m^2_{diq} + (\pi/D)^2}.
\ee

\begin{figure}
\centerline{
\epsfig{figure=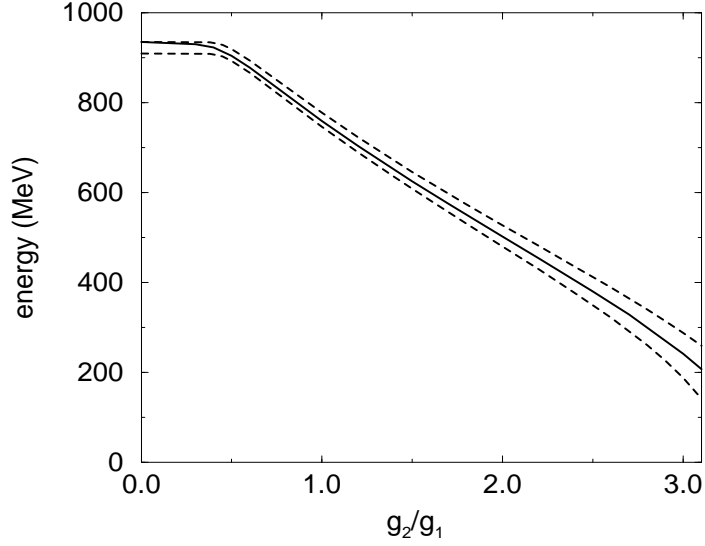,height=8.0cm,width=10.0cm}
}
\caption{\label{Kontivgl}
\sf The energy of the diquark in a spherical box (solid line) in
dependence of $g_2/g_1$, the dashed lines corresponds to the upper
($\omega_{diq}^{(1)}$) and lower ($\omega_{diq}^{(2)}$) limit of the
energy of the diquark.}
\end{figure}

As can be seen from Tab.~\ref{Diquarkmasse} and from
Fig.~\ref{Kontivgl} the diquark energy always lies within the two
boundaries (\ref{omega1}) and (\ref{omega2}). In fact, in the limit of
vanishing diquark coupling ($g_2\to 0$) the energy of the diquark
$\omega_{diq}$ is very well approximated by $\omega_{diq}^{(1)}$. The
difference at large coupling constants between $\omega_{diq}$ and
$\omega_{diq}^{(2)}$ appears because of numerical effects at the 
boundary where the wave functions do not vanish completely. In the
context of the hybrid model we consider bound states in the solitonic
background. In this case, the wave functions are localized at the
origin and the boundary effects are negligible.  

\newpage
\section{The baryon propagator}

The baryon propagator can be evaluated by comparison of
eq.~(\ref{A_b^2}) and (\ref{Nambu-Gorkov}) 
\be
 \frac{1}{2} \bar{\PPsi}^\alpha  (\GG^{val}_{B})_{\alpha \beta}
 \PPsi^\beta = \frac{1}{2} \bar{\xxi} \GG \xxi .
\ee
Using the definitions of $\xxi$ and $\bar{\xxi}$ (\ref{xxi}) we can
write 
\be
 (\GG^{val}_B)_{\alpha \beta} =
  \left( \begin{array}{cc} \Delta_\alpha G_{11} \Delta^{*}_\beta
                     & - \Delta_\alpha  G_{12}\Delta_\beta  \\
         -\Delta^{*}_\alpha G_{21} \Delta^{*}_\beta
              &  \Delta^{*}_\alpha G_{22} \Delta_\beta
                    \end{array} \right) ,
\ee
where we have used the inversion of the quark propagator 
\be
 \GG_0 &=& \left( \begin{array}{cc} G_{11} & G_{12} \\
         G_{21} & G_{22}  \end{array} \right)
\ee
which is obtained from eq.~(\ref{green}) by neglecting the baryon
sources on the diagonal elements. The elements of this matrix are
given by  
\be 
\label{G_11}
 G_{11}&=& \left( 1 + \G_0 \Delta \tilde{\G_0} 
             \tilde{\Delta} \right)^{-1} \G_0 ,\nonumber \\
 G_{12}&=& -\G_0 \Delta \left( 1 + \tilde{\G_0} \tilde{\Delta}
             \G_0 \Delta \right)^{-1} \tilde{\G_0} V , \nonumber\\
 G_{21}&=& -V^{-1} \tilde{\G_0}^{-1} \tilde{\Delta}
             \left( 1 + \G_0 \Delta \tilde{\G_0} 
             \tilde{\Delta} \right)^{-1} \G_0 , \\
 G_{22}&=& -V^{-1} \left( 1 + \tilde{\G_0} \tilde{\Delta}
             \G_0 \Delta \right)^{-1} \tilde{\G_0} V .\nonumber
\ee

\section{Calculation of static observables}

The calculation of observables in the hybrid model separates in a
valence quark part and a sea part:
\be
 \langle \hat{O} \rangle = \langle \hat{O} \rangle_{val} + \langle
 \hat{O} \rangle_{sea} .  
\ee
Because of the ansatz for the baryon wave function (\ref{ProdAnsatz})
\be
 | B \rangle = \sum_{\alpha}\sum_{\nu}^{(\epsilon_\nu>0)}
 a_{\alpha\nu} | \alpha , \nu \rangle 
\ee
with 
\be
 B(x,y) = \langle x,y  | B \rangle = \int d\omega dE 
 \sum_{\alpha}\sum_{\nu}^{(\epsilon_\nu>0)} 
 a_{\alpha\nu}(E,\omega) B_{\alpha\nu}^{0}(\vec{x},x_4;\vec{y},y_4)
\ee
we can in addition separate the valence quark part in a diquark and a
residual valence quark part: 
\be
\label{valence_op}
 \langle \hat{O} \rangle_{val} &=& \langle B | \hat{O} | B
 \rangle_{val} \\
 &=& \sum_{\alpha,\beta} \sum_{\nu,\mu}^{(\epsilon_{\nu,\mu}>0)}
 a_{\alpha\nu}^* a_{\beta\mu} \langle \alpha , \nu | \hat{O} | \beta ,
 \mu \rangle \nonumber \\
 &=& \sum_{\alpha,\beta} \sum_{\nu}^{(\epsilon_{\nu}>0)} a_{\alpha\nu}^*
 a_{\beta\nu} \langle \alpha | \hat{O} | \beta \rangle 
 + \sum_{\alpha} \sum_{\nu,\mu}^{(\epsilon_{\nu,\mu}>0)}
 a_{\alpha\nu}^* a_{\alpha\mu} \langle \nu | \hat{O} | \mu \rangle
 \nonumber \\ 
 &=& \langle \hat{O} \rangle_{diq} + \langle \hat{O} \rangle_{Fad} .
 \nonumber 
\ee
To take into account the internal quark structure of the diquarks we
use the form  
\be
\label{op}
 O &=& \langle \hat{O} \rangle = \int \D \qq \D \bar{\qq} \ q^\dagger
 \hat{O} q \exp \left( \frac{1}{2} \int \bar{\qq} \GG^{-1} \qq \right)
 \\ \nonumber 
 &=& \frac{1}{2} \int \D \qq \D \bar{\qq} \ \bar{\qq} \hat{\OO} \qq
 \exp \left( \frac{1}{2} \int \bar{\qq} \GG^{-1} \qq \right) \\
 \nonumber  &=&  \left. \frac{\partial}{\partial \beta}\int \D \qq \D
 \bar{\qq} \ \exp \left( \frac{1}{2} \int \bar{\qq} \left( \GG^{-1} +
 \beta \hat{\OO} \right) \qq \right) \right|_{\beta=0} \\ \nonumber
 &=& \left. \frac{\partial}{\partial \beta} \TTr \log \left( \GG^{-1} +
 \beta \hat{\OO} \right) \right|_{\beta=0} ,
\ee
where we have extended the operator $\hat{O}$ according to the
Nambu-Gorkov formalism:
\be
 \hat{\OO} = \left(\begin{array}{cc}
             \gamma_0 \hat{O}  &          0 \\
                   0           & -V \tilde{O} V^\dagger
            \end{array} \right)
\ee
with $\tilde{O}=V \hat{O}^T \gamma_0 V^{-1}$. Expanding the expression
(\ref{op}) in powers of the diquark field the zeroth order term
corresponds to $\langle \nu | \hat{O} | \mu \rangle$ in the quark term
whereas the second order term yields the diquark part $\langle \alpha
| \hat{O} | \beta \rangle$ in eq.~(\ref{valence_op}).  

The part of the residual valence quark is given by
\be
 \langle \hat{O} \rangle_{Fad} = \int \frac{d \tilde{\Omega}}{2\pi}
 \sum_{\alpha} \sum_{\nu,\mu}^{(\epsilon_{\nu,\mu}>0)}
 a_{\alpha\nu}^*(i\Omega_B,i\tilde{\Omega})
 a_{\alpha\mu}(i\Omega_B,i\tilde{\Omega}) \langle \nu | \hat{O} | \mu
 \rangle .
\ee

The sea part is calculated as usual in the pure soliton model
\cite{Alk96a}: 
\be
 \langle \hat{O} \rangle_{sea} = - \frac{1}{2} \sum_\nu \langle \nu |
 \hat{O} | \nu  \rangle \sgn(\epsilon_\nu)
 \rm{erfc} \left( \left| \frac{\epsilon_\nu}{\Lambda} \right| \right).
\ee

\subsection{Baryon number $\hat{B}$} 

In the Nambu-Gorkov formalism the baryon number operator is defined as 
\be
 \hat{\BB} = \left(\begin{array}{cc}
             \gamma_0 \hat{B}  &          0 \\
                   0           & -V \gamma_0 \hat{B} V^\dagger
            \end{array} \right) 
 \qquad \hbox{with} \qquad 
 \hat{B}=\ID_C/N_C=\ID_C/3.
\ee

The term of second order in the diquark field is expressed as 
\be
\label{B_diq}
 B_{diq} &=& i \int \frac{d\omega}{2\pi} dr^\prime r^{\prime 2}
 \Delta^*_\alpha(r^\prime,-i\omega) \int dr r^2 \Delta_\alpha(r,i\omega)
 \int \sum_{\nu
 \mu} \int d\Omega d\Omega^\prime
 R_2(\omega;\epsilon_{\nu},\epsilon_{\mu}) \\ \nonumber
 && \qquad \qquad \qquad \qquad \qquad \qquad * \left[
 \Psi_\mu^\dagger(\mbox {\boldmath $r^\prime $})
 \Psi_\nu(\mbox {\boldmath $r^\prime$}) \Psi_\nu^\dagger(\mbox
 {\boldmath $r $}) \Psi_\mu(\mbox {\boldmath $r $}) \right]
\ee
with the regularization function
\be
 R_2(i\omega;\epsilon_{\nu},\epsilon_{\mu}) &=& \\ 
 &&\frac{i\omega\left(\omega^2 + \left( \epsilon_{\nu} - \epsilon_{\mu}
 \right)^2 \right)}{4 \sqrt{\pi}} \int_0^1 d\alpha \alpha^2
 \frac{\Gamma\left(\frac{5}{2},\left[\alpha \epsilon_\mu^2 +
 (1-\alpha) \epsilon_\nu^2 +\alpha(1-\alpha)\omega^2 / \Lambda^2
 \right] \right)} {\left[\alpha \epsilon_\mu^2 + (1-\alpha)
 \epsilon_\nu^2 +\alpha(1-\alpha)\omega^2 \right]^{5/2}} \nonumber.
\ee

\subsection{Baryon density}

For the calculation of the baryon densities we need the identity
operator $\hat{O}=\ID$ as well as a spatial dependent variation
variable $\beta=\beta(\mbox {\boldmath $r$})$ in eq.~(\ref{op}). The
diquark contribution can be written as
\be
 B_{diq}(\mbox {\boldmath $r$}) = \sum_{\nu \mu}
 \Psi^\dagger_\nu(\mbox{\boldmath $r$}) \Psi_\mu(\mbox {\boldmath $r
 $}) R_3(i\omega;\epsilon_{\nu},\epsilon_{\mu}) 
\ee
with the regularization function
\be
\label{R3}
 R_3(i\omega;\epsilon_{\nu},\epsilon_{\mu}) &=&
 \frac{1}{4\sqrt{\pi}} \int_0^1 d\alpha \sum_\kappa \left[
 \frac{\Gamma\left(\frac{3}{2},\left[\alpha \epsilon_\mu^2 + 
 (1-\alpha) \epsilon_\kappa^2 +\alpha(1-\alpha)\omega^2 / \Lambda^2
 \right] \right)} {\left[\alpha \epsilon_\mu^2 + (1-\alpha)
 \epsilon_\kappa^2 + \alpha(1-\alpha)\omega^2 \right]^{3/2}} \right. \\
 && - \left. \frac{\Gamma\left(\frac{3}{2},\left[\alpha \epsilon_\nu^2 + 
 (1-\alpha) \epsilon_\kappa^2 +\alpha(1-\alpha)\omega^2 / \Lambda^2
 \right] \right)} {\left[\alpha \epsilon_\nu^2 + (1-\alpha)
 \epsilon_\kappa^2 + \alpha(1-\alpha)\omega^2 \right]^{3/2}} \right]
 \nonumber \\
 && \frac{[i\omega + \epsilon_\kappa - \epsilon_\nu][-i\omega +
 \epsilon_\kappa - \epsilon_\mu]}{\epsilon_\mu-\epsilon_\nu} \langle
 \mu |  \Gamma_{diq}(i\omega) | \kappa \rangle \langle \kappa |
 \tilde{\Gamma}_{diq}(-i\omega) | \nu \rangle \nonumber \\
 && \left( \epsilon_\mu+\epsilon_\nu \right)
 \frac{\Gamma\left(\frac{3}{2},\left[\alpha \epsilon_\mu^2 +  
 (1-\alpha) \epsilon_\nu^2 +\alpha(1-\alpha)\omega^2 / \Lambda^2
 \right] \right)} {\left[\alpha \epsilon_\mu^2 + (1-\alpha)
 \epsilon_\nu^2 + \alpha(1-\alpha)\omega^2 \right]^{3/2}} \langle
 \mu |  \Gamma_{diq} \tilde{\Gamma}_{diq}| \mu
 \rangle \nonumber 
\ee

\subsection{Axial coupling constant}

Using the operator $\hat{O}=-\frac{1}{3}\sigma^3\tau^3$ \cite{Alk96a}
the axial coupling constant is given by
\be
\label{g^diq_A}
 g^{diq}_A = -\frac{1}{3} \sum_{\nu,\mu} \langle \nu | \sigma^3\tau^3 |
 \mu \rangle R_3(i\omega;\epsilon_{\nu},\epsilon_{\mu}) 
\ee   
with the regularization function $R_3$ defined in (\ref{R3}). Note,
that eq.(\ref{g^diq_A}) is only valid for grand spin eigenstates
$| \mu \rangle$.

\end{document}